# Exploring contrast generalisation in deep learning-based brain MRI-to-CT synthesis

Lotte Nijskens[a,b], Cornelis A.T. van den Berg[a,b], Joost J.C. Verhoeff[b], Matteo Maspero[a,b,*]

[a]*Computational Imaging Group for MR Diagnostics & Therapy, Center for Image Science, University Medical Center Utrecht, Heidelberglaan 100, Utrecht, 3584CX, The Netherlands*
[b]*Department of Radiotherapy, University Medical Center Utrecht, Heidelberglaan 100, Utrecht, 3584CX, The Netherlands*

**Abstract**

**Background:** Synthetic computed tomography (sCT) has been proposed and increasingly clinically adopted to enable magnetic resonance imaging (MRI)-based radiotherapy. Deep learning (DL) has recently demonstrated the ability to generate accurate sCT from fixed MRI acquisitions. However, MRI protocols may change over time or differ between centres resulting in low-quality sCT due to poor model generalisation.

**Purpose:** investigating domain randomisation (DR) to increase the generalisation of a DL model for brain sCT generation.

**Methods:** CT and corresponding $T_1$-weighted MRI with/without contrast, $T_2$-weighted, and FLAIR MRI from 95 patients undergoing RT were collected, considering FLAIR the unseen sequence where to investigate generalisation.

A "Baseline" generative adversarial network was trained with/without the FLAIR sequence to test how a model performs without DR. Image similarity and accuracy of sCT-based dose plans were assessed against CT to select the best-performing DR approach against the Baseline.

**Results:** The Baseline model had the poorest performance on FLAIR, with mean absolute error (MAE)=106±20.7 HU (mean ±$\sigma$). Performance on FLAIR significantly improved for the DR model with MAE=99.0±14.9 HU, but still inferior to the performance of the Baseline+FLAIR model (MAE=72.6±10.1 HU). Similarly, an improvement in $\gamma$-pass rate was obtained for DR vs Baseline.

**Conclusions:** DR improved image similarity and dose accuracy on the unseen sequence compared to training only on acquired MRI. DR makes the model more robust, reducing the need for re-training when applying a model on sequences unseen and unavailable for retraining.

*Keywords:* medical imaging, radiotherapy, artificial intelligence, machine learning, regression, magnetic resonance imaging, computed tomography, generalization, domain shift

[*]Corresponding author
*Email address:* m.maspero@umcutrecht.nl (Matteo Maspero)
*URL:* https://orcid.org/0000-0003-0347-3375 (Matteo Maspero)

# 1. Introduction

Radiation therapy (RT) is one of the main pillars of cancer treatment, indicated for approximately half of all cancer patients [1]. Computed tomography (CT) is the primary imaging modality for RT planning, providing geometrically accurate electron density required for dose calculations[2]. In parallel, magnetic resonance imaging (MRI) has superior soft tissue contrast compared to CT. Hence, MRI has been suggested as the preferred imaging modality for delineating tumours and the surrounding OARs[3]. Acquiring MRI can significantly reduce the intra- and interobserver variability in the tumours and OARs delineations for multiple disease sites, including the breast [4], prostate, and head-and-neck [5]. Additionally, for certain brain cancer patients, MRI is key to resolve tumour boundaries unidentified on CT[6, 7].

Unlike CT, MRI does not inherently contain the information needed for dose calculations[8]. Whenever MRI provides additional information, CT and MRI are both acquired and fused for RT planning[9], possibly introducing systematic uncertainties due to residual misregistrations[10, 11].

MRI-only based RT has been proposed to avoid residual error after image registration in a hybrid CT-MRI workflow[12, 13]. Also, MRI-only RT reduces the patient's exposure to ionising radiation[9], which significantly benefits when re-planning is needed[14] or for paediatric populations[15]. Another advantage is that fewer scans are required, improving patient comfort[16]. Moreover, eliminating the CT scan and simplifying the workflow reduces the workload[17, 18] and costs[19, 17, 16]. Finally, with the clinical introduction of MRI-guided RT[20, 21], MRI-only RT became particularly interesting[22, 23].

The lack of a physical relationship between the tissues' nuclear magnetic properties and their electron density characteristics needed for dose calculations can be regarded as the main hurdle in implementing MRI-only RT. Many approaches for representing MRI as a CT equivalent have been proposed to overcome this problem[24] obtaining the so-called synthetic CT (sCT). Alternative names for the resulting image are pseudo-CT, MR-CT, virtual CT, or substitute CT[17, 25].

Most recently, deep learning (DL)-based methods became of particular interest as their inference requires limited time (seconds to minutes) for sCT generation, unlike classical image processing-based methods (ten minutes to hours)[26, 27]. This time aspect is essential in MRI-guided RT, requiring sCT generation within a few minutes to allow daily re-planning[25, 22].

Convolutional neural networks (CNNs) are the most common and most successful DL models applied to (medical) imaging[28], achieving convincing performances for image synthesis tasks[24]. Even if DL is a promising approach, DL models are known to generalise to new domains poorly[29, 30, 31]. Models assume a shared statistical distribution and feature space between the training and test data, meaning they must be re-trained if the test data lies outside the distribution[29]. For sCT, new domains could be, e.g., a different MRI sequence, an MRI acquired in another hospital with varying parameters of acquisition, MRI received after a scanner upgrade or with a different model, or another anatomy.

Most DL models for sCT generation presented so far are trained and tested on a specific anatomical site using MRI acquired with limited sequences and a fixed range of imaging parameters[24]. These models do not consider the variability in MRI acquisition protocols employed in clinical practice or protocol changes that might occur over time. Ideally, robust and general models that can output sCT from MRI sequences unseen during training would facilitate a trustworthy and widespread clinical implementation[32, 33]. However, this is one of the challenges deep learning techniques still need to face[34, 35].

Two recent publications investigated generalisation to multiple MRI sequences in MRI-only RT[36, 37]. The methods employed in these works to improve inference performances involved (re)training the network on those sequences. Zimmermann *et al.*[37] showed that it is possible to train a combined model for T1w(Gd) and T2w images. A comparison with single-sequence models revealed poor generalisation to the sequence not included in the training data[37], as also found by Li *et al.*[36].

Recently, a promising technique called domain randomisation was proposed to improve a segmentation network's ability to generalise to unseen MRI sequences[38, 39]. The method relies on the hypothesis that increasing variability in synthetic training data forces the model to provide accurate output for all domains, e.g., in MRI sequences[40] and is motivated by experiments showing that data augmentation beyond realism improves generalisation[41].

This work investigates the use of domain randomisation to obtain DL-based models for MRI-to-sCT generation that can generalise to MRI scans acquired with unseen sequences in brain MRI-only RT. Inspired by Billot *et al.*[38, 39], we propose a domain randomisation method that converts the MRI into synthetic images of random contrast to improve contrast generalisation. The hypothesis is that training a DL model for MRI-to-sCT generation on input data with synthetic, not necessarily realistic, image contrasts obliges the network to learn contrast-agnostic features[40, 41].

Two approaches to domain randomisation are investigated: 1) training on images with *random contrast* synthetically generated from segmented MRIs and 2) training on random *linear combination*s of multiple MRI sequences. The effects of domain randomisation are compared to training solely on a mix of acquired sequences. The end goal is to investigate how far a DL model can be pushed towards becoming contrast-agnostic, capable of generating sCT, from which RT plans can be calculated with clinically acceptable dose accuracy.

## 2. Materials and methods

### 2.1. Data collection and imaging protocols

Data from 95 patients were selected undergoing treatment at the UMC Utrecht RT department from a large retrospective, anonymised database collected under the local Medical Ethical Committee's approval (study number: 20/519, approved on August 11, 2020). The main selection criterion was the availability of a treatment plan for brain RT conducted between January 2020 and July 2021, with corresponding CT and MRI ($T_1$-weighted with and without contrast enhancement, $T_2$-weighted and FLAIR images). Patients were excluded if not all sequences were available, no suitable CT was available, the time between MRI and CT acquisition was more than 1.5 months, the patient's age was < 18 years, or the MRI was a follow-up exam. If multiple CT acquisitions were available, the most recent one was chosen, and the MRI dataset acquired closest in time to the CT was selected.

Patients were randomly divided over the training (n=60), validation (n=10) and test set (n=25). The female/male ratio for the 95 included patients was 51/44 with a mean patient age of 59.9 ±13.0 years (range: 24.3-86.8). The mean interval between CT and MRI acquisition was six days (range: min-max =1-26). Dose prescriptions ranged from 14 to 60 Gy over 1-33 fractions.

Planning CTs were acquired at the radiotherapy department using a Brilliance Big Bore system (Philips Healthcare, USA). The acquisition occurred in the supine treatment position, aided by head support and a personalised immobilisation mask. CT acquisition was without contrast agents, with a tube potential of 120 kV, a tube current of 234-360 mA (range=min-max), and 1000-1712 ms exposure. The in-plane resolution was 0.57-1.17 mm$^2$, with a slice thickness of 1-2 mm.

MRI data were acquired with a 1.5 or 3.0 T Ingenia MR-RT system (Philips Healthcare, the Netherlands). Available sequences (Table 1) were: 3D $T_1$-weighted turbo field echo (TFE) images with and without Gadolinium contrast (T1w and T1wGd), 2D $T_2$-weighted turbo spin-echo (TSE) images with Gadolinium contrast (T2w) and 3D $T_2$-weighted FLAIR TSE images (FLAIR).

Table 1: Overview of acquisition parameters per sequence for the 95 included patients.

| Parameter | 3D T1w TFE | 3D T1w TFE Gd | 2D T2w TSE | 3D T2w FLAIR TSE |
|---|---|---|---|---|
| $B_0$ [T] | **1.5** (66) | **1.5** (66) | **1.5** (66) | **1.5** (66) |
|  | **3.0** (29) | **3.0** (29) | **3.0** (29) | **3.0** (29) |
| Contrast | No | Yes | Yes | No |
| Read-out | AP | AP | AP | AP |
| Flip angle [°] | 8 | 8 | 90 | 90 |
| TR [ms] | 7.6-8.7 | 7.6-8.7 | 3119-5996 | 4800 |
| TE [ms] | 3.5-4.1 | 3.5-4.1 | 80-100 | 303-363 |
| FOV[a] [mm$^3$] | 230, 160 | 230, 160 | 230, 140-160 | 230, 160 |
| Acq voxel[a] [mm$^3$] | 1.0, 0.5-1.0 | 1.0, 0.5-1.0 | 0.6-0.7, 4.0-5.0 | 1.1-1.2, 0.6 |
| Rec voxel[a] [mm$^3$] | 0.4-1.0, 0.5-1.0 | 0.5-1.0, 0.5-1.0 | 0.4-0.5, 4.0-5.0 | 1.0, 0.6 |
| Rec matrix[a] | 240-512,162-323 | 240-480,162-323 | 480-512,31-43 | 240,269-270 |
| BW [Hz/px] | 190-217 | 190-217 | 143-206 | 851-1075 |
| Acq time [s] | 136-271 | 121-271 | 117-137 | 331-475 |

Acq time: acquisition time; Acq voxel / Rec voxel: acquisition/reconstruction voxel size; AP: anterior-posterior; $B_0$: main magnetic field strength; BW: bandwidth; FOV: field-of-view; Rec matrix: reconstruction matrix; TE: echo time; TFE: turbo field echo; TR: repetition time; TSE: turbo spin echo. [a]Directions: anterior-posterior, right-left and craniocaudal.

## 2.2. Image processing

If not otherwise specified, image processing and performance evaluation was performed in Matlab R2019a (The MathWorks, Inc., USA).

**Pre-processing** Each MRI was rigidly registered to the corresponding CT with `Elastix` (version 4.700)[42, 43], using multi-resolution registration (with a resolution of 4, 2, 1, and 0.5 times the reconstructed voxel size) with an adaptive stochastic gradient descent optimiser and mutual information similarity metric. The parameters from[44] were adopted. The registered MRI will be referred to as $MRI_{reg}$. $MRI_{reg}$ and CT were resampled to isotropic 1.0x1.0x1.0 mm$^3$ resolution using linear interpolation.

Most images contained a discrepancy between CT and MRI FOVs, caused by angled MRI acquisition (Fig. 1). A body mask was computed on the non-registered MRI to ensure congruent FOVs between CT and $MRI_{reg}$. The mask was generated using a threshold with an empirically determined value of 20 (or 15 for T2w images), followed by morphological filling and dilation with a disk-shaped structuring element of radius 20 voxels. The binary mask was registered to the CT by applying the transform computed for $MRI_{reg}$, resampled, and applied to the $MRI_{reg}$-CT pair for training. The $MRI_{reg}$ and CT FOVs were cropped to the extent of the registered mask with additional ten voxel margins on each side or until the original image boundary. $MRI_{reg}$ were normalised by clipping to the per-patient 99th percentile over the masked volume. Training CTs were clipped to range [-1024, 1500] HU.

For CT, the masking and range clipping steps were only applied to the training images (hereafter: $CT_{train}$). $CT_{train}$ and normalised $MRI_{reg}$ were saved as 3D volumes in NifTI format, linearly rescaled to [-1, 1]. Fig. 1 shows an example of a normalised brain $MRI_{reg}$ with the corresponding normalised $CT_{train}$, ground truth $CT_{crop}$ and original, unregistered MRI for each sequence for a single patient.

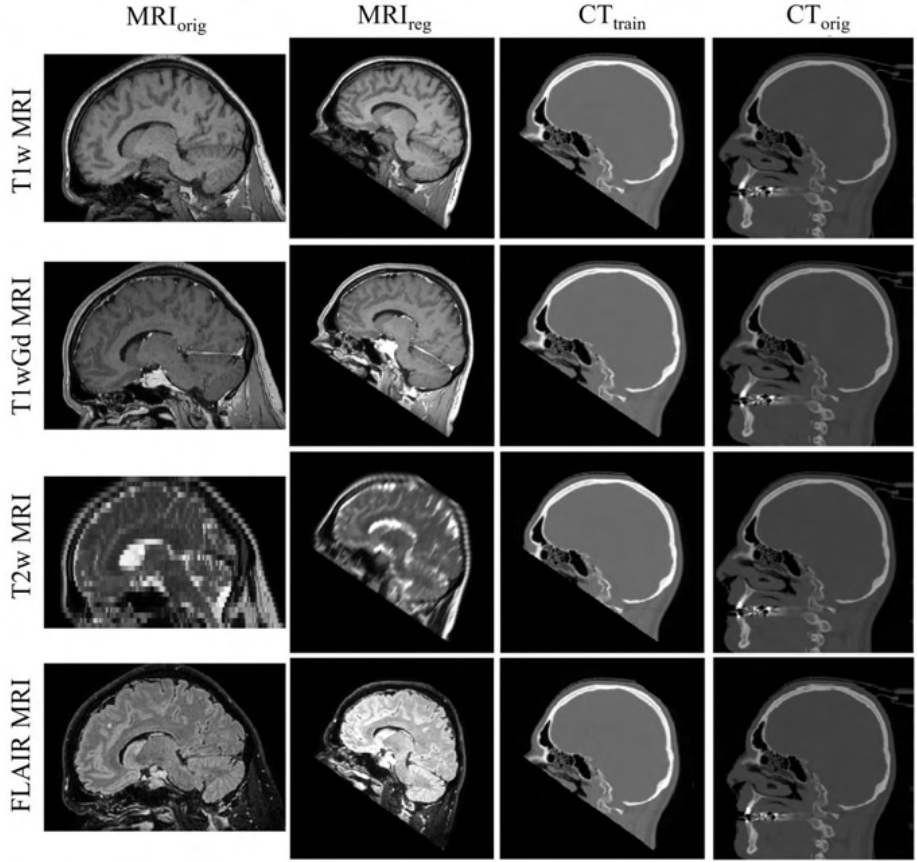

Figure 1: Example of the pre-processing outcomes. The original T1w (top row), T1wGd (second row), T2w (third row) and FLAIR (bottom row) brain MRI for a single male patient in the training dataset are shown (left) with corresponding normalised $MRI_{reg}$, $CT_{train}$ and ground truth $CT_{crop}$ (left to right).

**Post-processing** For post-processing all generated sCTs were linearly rescaled to a [-1024, 1500] HU range, conforming to the range of $CT_{train}$.

### 2.3. Performance evaluation

The quality of the generated sCTs was evaluated in terms of image similarity between acquired CT and generated sCT and dose accuracy. Statistical comparisons were performed with Wilcoxon signed-rank tests, with p-values < 0.05 regarded as statistically significant. Also, training and inference times are reported.

#### 2.3.1. Image similarity

The accuracy of the assigned HU values was analysed with a voxel-wise comparison between $CT_{crop}$ (ground truth) and sCT. A body contour mask was applied to $CT_{crop}$ and sCT before calculating the metrics and comparing on the intersection of the two masks. The masks were created by thresholding the (s)CT above -200 HU, then morphologically closing and filling the combined mask to include the nasal cavities. The mean absolute error (MAE) was computed per patient. Peak signal-to-noise ratio (PSNR) and structural similarity index measure (SSIM) were additional metrics. The range and mean ± standard deviation ($\mu \pm \sigma$) over all patients in the validation or test set were calculated for each metric.

*2.3.2. Dose accuracy*

The clinically optimised dose plan was re-calculated for the final models on (s)CT. Generated sCTs were registered and resampled to the original, non-cropped CT, allowing only translations. Multi-resolution registration was performed (with a factor 4, 2, 1, and 0.5 to the voxel size) with an adaptive stochastic gradient descent optimiser and mutual information. In case the registration quality was considered poor by an observer or failed, three resolutions were used instead of four (resolution 1: 4, 4, 2; resolution 2: 2, 2, 1; resolution 3: 1, 1, 0.5).

A segmentation of the body contour of the non-cropped CT was taken from the clinical treatment plan, and the voxels outside the original MRI FOV and inside this body contour were set to 0 HU. The difference in FOV between sCT and acquired CT was thus water-filled in both images. The water-filled sCT and acquired CT are referred to as $sCT_{wf}$ and $CT_{wf}$.

Plans were volumetric modulated arc therapy (VMAT) photon plans with a single arc with a beam energy of 6.0 MV. They were calculated with a Monte Carlo algorithm on a 3 mm$^3$ grid with 3% uncertainty. Plan re-calculation was performed on (s)$CT_{wf}$ using GPUMCD[45].

Dose accuracy was assessed through the calculation of the dose difference (DD) relative to the prescribed dose ($D_{presc}$) in the high-dose region (D > 90% of $D_{presc}$)[24]:

$$DD = 100 * \frac{D_{CT} - D_{sCT}}{D_{presc}}\%, \tag{1}$$

with D the dose (in Gy) in the (s)$CT_{wf}$-based dose plan. Korsholm *et al.*[46] proposed a criterion for the clinical acceptability of DD: the DD compared to a CT-based dose plan should be <2% for 95% of the patients. In this work, a more conservative criterion was adopted. Individual sCTs were considered acceptable if the DD was <1%.

Dose-volume histograms (DVH) were analysed for differences in $D_{median}$ and $D_{max}$ between sCT- and CT-based plans for the following OARs: brainstem, optic chiasm, lenses, cochleae, and pituitary gland. Additionally, a 3D-$\gamma$ global analysis was conducted[47]. For the computation of $\gamma$-pass rates, a 10% dose threshold was used, with 3%,3mm, 2%,2mm, and 1%,1mm criteria. Heilemann *et al.*[48] demonstrated the ability to detect clinically unacceptable VMAT plans using a 90% $\gamma$-pass rate threshold for the 2%,2mm criterion. Nevertheless, the absence of clinically significant dose differences was not guaranteed[48]. Considering that evaluation of $\gamma$-pass rates is adopted for quality assurance of delivered plans, where uncertainty is higher, we adopted stricter thresholds in this work: 95% and 99% for the 2%,2mm, and 3%,3mm criteria. The primary metric was the 95% dose threshold on $\gamma$ 2%,2mm.

*2.4. Network architecture*

The cGAN model `pix2pix` was implemented to allow paired training, as proposed by Isola *et al.* [49]. Initial investigations showed that 3D models outperformed 2D ones[50]. Therefore, only 3D models are reported in this work.

An implementation of the original `pix2pix` model [49] called `Ganslate` [51] in PyTorch version 1.10 was used for 3D models. All models were trained using a GPU Tesla P100 PCIe 16 GB or V100 PCIe 32 GB (NVIDIA Corp., USA).

A 3D U-Net generator architecture that allows variable patch sizes as input was adopted, along with a 70 x 70 PatchGAN discriminator [49]. The $L$1-based loss function proposed in [49] was implemented.

*2.5. Model optimisation*

Hyperparameter optimisation was performed with a grid search strategy on a subset of the training set consisting of ten patients from the training set with only T1w images without contrast. The hyperparameters leading to the lowest average MAE in the validation set were adopted. SSIM and PSNR were calculated as additional metrics. The hyperparameter grid search space is detailed in Supplementary Material I.A.

One patient was retrospectively excluded from the validation set after observing the registration failure between the T2w MRI and the CT. Validation of all models except those trained in the hyperparameter tuning stage was thus done on a nine-patient validation set.

As a final optimisation step, the ratio between T1w images with/without contrast and T2w images in the training set was balanced, and the batch size was finetuned (Supplementary Material I.B. ). This step involved training a subset of fifteen patients from the training set. After balancing, the final training dataset (n=60) contained 60 T2w, 30 T1w, and 30 T1wGd images.

After optimisation, all models were trained with Xavier initialisation, Adam optimiser, patch size = 128x128x128 voxels, batch size = 1, $\lambda$ = 5000, number of downsampling steps = 5, and a constant learning rate of 0.001. A sliding window was used for patch combination with a patch overlap of 0.5 and Gaussian blend mode. The Adam optimiser [52] was implemented with $\beta_1$ = 0.5 and $\beta_2$ = 0.999 as momentum parameters and no weight decay.

Early stopping was applied to avoid overfitting, using the MAE as a decision criterion. The MAE was calculated for the sequences included for training, in the body contours for the patients in the validation set. A combined MAE was computed as the average for the seen sequences. The first iteration for which this combined MAE did not improve for the following three iterations was selected, evaluating every 50,000 iterations. Supplementary Material I.C. illustrates the early stopping method.

*2.6. Domain randomisation*

Domain randomisation was applied by training on generating random contrast (RC) images starting from label maps or generating linear combinations (LC) of the acquired MRI. Both approaches have been investigated and are described in the following.

*2.6.1. Random contrast*

The domain randomisation strategy comprising RC images (section 2.6) requires segmenting patients' MRIs. Automatic segmentations were generated from the T1w images, complemented by some structures labelled using $CT_{train}$.

Segmentation of cerebral structures was performed on T1w images using an open-source DL network called FastSurfer [53]. OARs were added by segmentation of T1w MRI with a previously in-house developed DL algorithm that is clinically adopted (unpublished and developed for clinical use as in [54]), based on the DeepMedic model [55]. The GTV was obtained from the clinical segmentation. Cerebrospinal fluid (CSF) was labelled using a combination of FastSurfer labels and clinical segmentations. Also, $CT_{train}$ was segmented by thresholding to obtain labels for bone and soft tissue. The resulting label maps were stored as an additional dataset. More details on image segmentation, a lookup table with included labels, and an example of created label maps are reported in Supplementary Material II.

Label maps were converted to RC images for network training, as proposed by Billot et al. [39] using TorchIO library [56]. Specifically, after randomly selecting a segmentation from the training data, each label was assigned a Gaussian function with mean and standard deviation chosen randomly from a uniform distribution with ranges of [10, 240] and [1, 25], respectively. These ranges were based on the sensitivity analysis conducted in [39]. All voxels within a label were assigned an intensity value sampled from this Gaussian distribution.

Then, images were blurred to increase spatial coherence between neighbouring voxels. The standard deviation of the Gaussian was randomly sampled from a uniform distribution: $\sigma_{blur} \sim U(0, 0.3)$, like in [38]. Random gamma augmentation was applied after rescaling image intensity to a positive range to increase variability in the training data further. Following [39], the exponent $\gamma = e^\beta$ was randomly sampled from a normal distribution: $\beta \sim N(\mu_\beta, \sigma_\beta)$ with $\mu_\beta = 0$ and $\sigma_\beta = 0.4$. The RC image was rescaled to [-1, 1]; see Fig. 2(top) for an example of RC image patches.

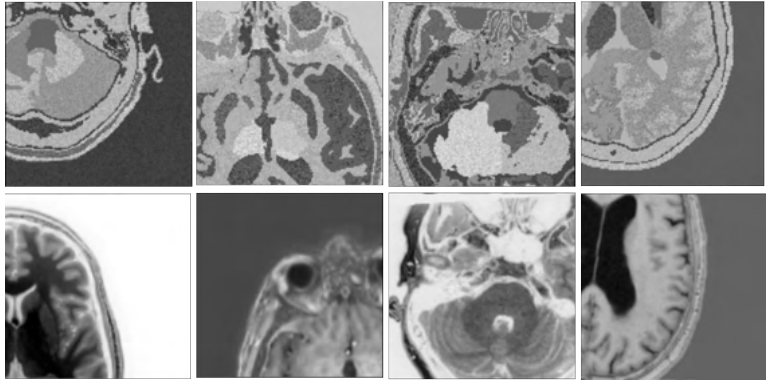

Figure 2: Examples of random contrast (RC; top row) and linear combination (LC; bottom row) images generated from label maps. Each image is a slice from an example patch as input to the network during training.

*2.6.2. Linear combination*

Linear combination was considered to investigate a more straightforward method to perform domain randomisation than RC that would not require brain segmentations, facilitating the application of domain randomisation.

Linear combination (LC) images were generated from T1w(Gd) and T2w MRI. To enable linear combination, an additional training dataset was created for which the pre-processed T2w and T1wGd images were registered to the corresponding T1w image with the same registration parameters as in section 2.2 A random choice was made during network training between combining the patient's T2w image with their T1w or T1wGd image, using equal probabilities. The T1w(Gd) and T2w images were then combined as follows:

$$Im_{LC} = p_1 * Im_{T1} + p_2 * Im_{T2}, \qquad (2)$$

with $p_1$ and $p_2$ as the coefficients for voxel-wise addition of the T1w(Gd) ($Im_{T1}$) and T2w ($Im_{T2}$) image, respectively. These were randomly sampled from a uniform distribution: $p_{1,2} \sim U(-1, 1)$. The chosen range allows addition and subtraction in the linear combination and contrast inversions. Finally, images were rescaled to the range [-1, 1]. Figure 2(bottom) shows several examples of LC patches.

*2.7. Experiment: random contrast vs linear combinations*

An experiment was conducted to identify the most effective domain randomisation technique between RC and LC, comparing two models in terms of image similarity on the validation set. Each model was trained with both the two domain randomisation approaches.

A model trained on a mix of acquired MRI and RC images derived from segmentations was adopted for this experiment to represent the RC method after initial investigations showed that it outperformed a model trained on RC images only [50]. For model training, the entire training dataset was used. Hence, this RC model was trained on 60 segmentations, 30 T1w, 30 T1wGd, and 60 T2w images from the training set (n=60 individual patients).

For the domain randomisation method comprising LC images, initial investigations showed that a model trained with a 50% chance of applying a linear combination to the acquired MRI outperformed a model trained with a 100% chance of using a linear combination [50]. The LC model was thus trained with a 50% chance of linear combination. The dataset from which LC images were generated consisted of acquired T1w (n=60) images and T1wGd (n=60) and T2w (n=60) images that had been registered to their T1w counterpart, as mentioned in section 2.6.2 For the LC model, this LC-specific dataset and the original training dataset of 30 T1w images, 30 T1wGd images and 60 T2w images (section 2.5) were used. At each iteration, a random choice was made whether to apply a linear combination. The original dataset was sampled if an LC image should not be used.

The RC and the LC model were trained with the hyperparameters described in section 2.5. Early stopping was based on the MAE obtained for sCT generated from T1w(Gd) and T2w images in the validation set for both models.

The RC and LC models were statistically compared using image similarity metrics calculated per sequence (T1w(Gd), T2w and FLAIR) on the validation set using MAE as the leading metric for model choice. The best-performing model was adopted as the final Domain Randomisation model.

*2.8. Domain randomisation on an unseen sequence*

In a final comparison, the chosen Domain Randomisation model that will result from the experiment described in 2.7 was compared to two models trained without domain randomisation: a Baseline model and a Baseline+FLAIR model. In this way, we can compare the impact of adding the unseen FLAIR sequence against domain randomisation compared to the baselines.

*2.8.1. Baseline model*

The Baseline model was trained on a mix of T1w, T1wGd and T2w images to assess the models' ability to generalise to an unseen sequence (FLAIR) without domain randomisation. For model training, the entire training dataset was used: 30 T1w, 30 T1wGd, and 60 T2w images, applying the hyperparameters described in section 2.5. The early stopping iteration was determined based on the MAE on T1w, T1wGd and T2w images of patients in the validation set.

After determining when to apply early stopping on the valiation set, the model was inferred on the test set (n=25). Image similarity metrics and dose accuracy were calculated for sCT generated from patients' T1w, T1wGd, T2w and FLAIR images in the test set. The image similarity metrics and metrics for dose evaluation were statistically compared between the four sequences. For the discussion of the dose accuracy on an individual patient level, we limit ourselves to the results obtained for the unseen FLAIR sequence.

*2.8.2. Baseline+FLAIR vs Baseline*

The Baseline+FLAIR model was trained to obtain a measure for the best achievable performance for FLAIR input images. This model was trained on the whole training set of 60 patients used for training the Baseline model, with the addition of FLAIR images. Hence, altogether the training dataset consisted of a mix of T1w (n=30), T1wGd (n=30), T2w images (n=60) and FLAIR images (n=60) from 60 individual patients. The hyperparameters described in section 2.5 were adopted. For the Baseline+FLAIR model, the iteration for early stopping was determined based on the MAE on T1w, T1wGd, T2w and FLAIR images of patients in the validation set.

After early stopping, the Baseline+FLAIR model was inferred on the test set (n=25). Image similarity metrics and dose accuracy were calculated for sCT generated from each of the four sequences (T1w, T1wGd, T2w and FLAIR). For each sequence, the image similarity metrics and metrics for dose evaluation were statistically compared with those obtained for the Baseline model. Also, statistical comparisons were made between the sequences.

*2.8.3. Domain Randomisation vs baselines*

For the final comparison, the Domain Randomisation model was inferred on the test set (n=25 patients). As for the Baseline and Baseline+FLAIR model, image similarity metrics and dose accuracy were calculated for sCT generated from patients' T1w, T1wGd, T2w and FLAIR images. Per sequence, the image similarity and dose evaluation metrics were statistically compared to those obtained for the Baseline and Baseline+FLAIR model. Additionally, statistical comparisons were made between the sequences.

# 3. Results

For all the models inference time on the test set was approximately 4 s per sequence and patient.

*3.1. Experiment: random contrast vs linear combinations*

Early stopping was applied after 450,000 iterations for the RC model and 200,000 iterations for the LC model. For all sequences, the MAE obtained on the validation set was lower for the RC model than for the LC model (Table 2). Only the difference in FLAIR images was statistically significant: an MAE of 105±20.5 HU was obtained for the RC model, compared to an MAE of 110±23.9 HU for the LC model. Differences in SSIM and PSNR were not statistically significant (Supplementary Material III.A.).

Overall, using RC images was deemed the most beneficial domain randomisation strategy. Consequently, the RC model was adopted as the Domain Randomisation model for final comparison with the Baseline and Baseline+FLAIR models.

Table 2: MAE obtained for sCT generated by the RC and LC models per MRI sequence. Metrics were calculated on the validation set (n = 9) within the intersection of the body contour of the sCT and CT. Mean values and standard deviations ($\mu \pm 1\sigma$) and range ([min - max]) are reported. Wilcoxon-signed rank tests were used for statistical comparisons. Values of p < 0.05 were regarded as statistically significant.

| Metric | Model | Sequence | | | |
|---|---|---|---|---|---|
| | | T1w | T1wGd | T2w | FLAIR |
| MAE [HU] | RC | 71.5±12.1 [59.7 - 100] | 69.6±12.2 [56.6 - 98.6] | 76.3±10.9 [60.2 - 95.6] | 105±20.5 [74.1 - 142] |
| | LC | 72.3±12.4 [57.3 - 100] | 71.0±12.2 [58.2 - 99.8] | 77.8±11.4 [63.0 - 100] | 110±23.9 [72.9 - 155] |
| p-value | | 0.3 | 0.2 | 0.2 | 0.04 |

### 3.2. Domain randomisation on an unseen sequence

#### 3.2.1. Baseline model

The training time for the Baseline model was 32.0 h, applying early stopping at iteration 300,000.

Among the four sequences, the performance of the Baseline model on the test set (n=25) was best on T1w and T1wGd images (Fig. 3), with the difference between these two sequences not statistically significant for the three metrics. The p-values resulting from statistical tests of performance metrics (image similarity and dosimetric accuracy) between sequences are presented in Supplementary Material III.B. for each of the models. The mean MAE was 64.2±7.3 HU or 63.8±9.1 HU for T1w and T1wGd images, respectively. The worst performance was found for FLAIR images, with a considerable difference with performance on T1w(Gd) images: the mean MAE was 106±20.7 HU. Testing on T2w images resulted in a mean MAE of 69.6±8.5 HU. The difference in performance on FLAIR and T2w images compared to the performance on the other three sequences was statistically significant. Violin plots for SSIM and PSNR are shown in Supplementary Material III.C. Results for SSIM and PSNR were in line with the MAE for the Baseline model.

For the Baseline model, 3D $\gamma$-pass rates in the low dose region (> 10% of the prescribed dose) with 1%,1mm criterion were > 95% (Table 3) for every patient and each sequence. The pass rate obtained for FLAIR images (99.0±1.1%) was significantly lower than that computed for all other sequences. The $\gamma$-pass rates with 3%,3mm and 2%,2mm criteria were >99% for every patient and each sequence. The obtained $\gamma$-pass rates with 3%,3mm and 2%,2mm criteria are shown in Supplementary Material III.D. for each of the models.

For the Baseline model, a DD in the high-dose region (> 90% of the prescription dose) of -0.1±0.2% was obtained for treatment plans based on sCT generated from T1w, T1wGd and T2w images, and a DD of 0.4±0.5% was found for FLAIR images (Table 3). The DD was significantly larger for FLAIR than the three other sequences. Other differences in DD between sequences were not statistically significant.

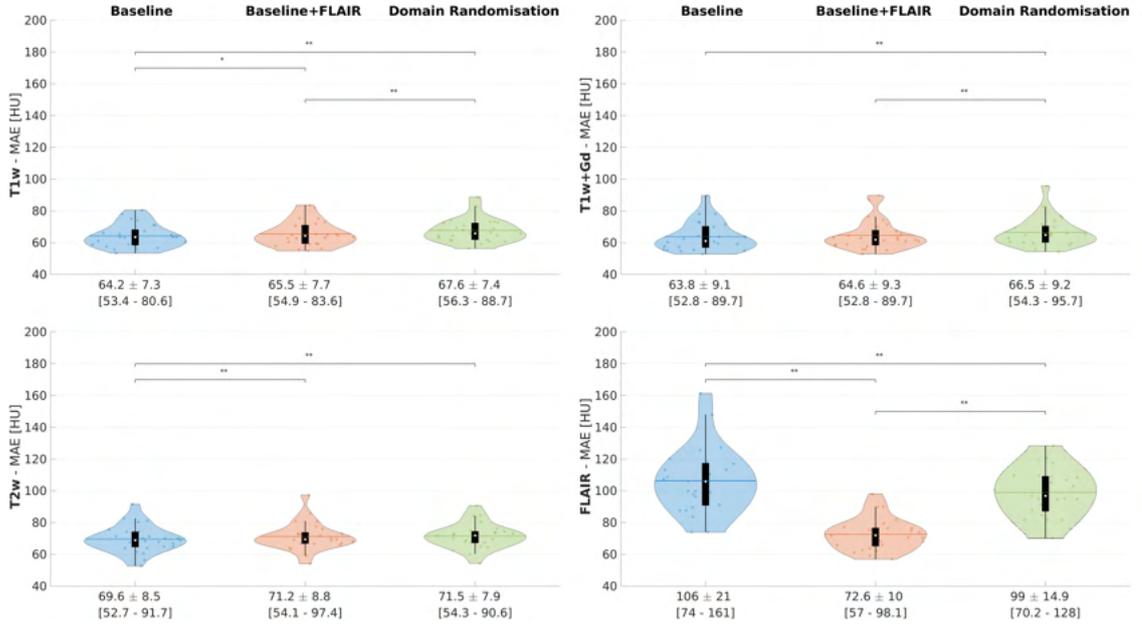

Figure 3: Violin and box-and-whisker plots for the MAE in the intersection of the body contour of sCT compared to ground truth CT on the test set (n=25) for sCT generated by the Baseline (blue), Baseline+FLAIR (orange) and Domain Randomisation model (green). Results are presented per sequence: T1w (top left), T1wGd (top right), T2w (bottom left) and FLAIR (bottom right). The black box indicates the interquartile range and median (white circle) with whiskers indicating the range, outliers excluded. The width of the violin indicates the distribution of the data points. The mean values and standard deviations are shown. Statistically significant differences are indicated by * (p< 0.05) or ** (p< 0.001). Wilcoxon-signed rank tests were used for statistical testing.

Table 3: Dose evaluation ($\gamma_{1\%,1mm}$ and DD) for sCT generated by the Baseline, Baseline+FLAIR and Domain Randomisation models per MRI sequence. Dose accuracy was assessed through plan re-calculation on sCT$_{wf}$ compared to CT$_{wf}$. Mean values and standard deviations ($\mu \pm 1\sigma$) and range ([min - max]) are reported.

| Metric | Model | Sequence | | | |
|---|---|---|---|---|---|
| | | T1w | T1wGd | T2w | FLAIR |
| $\gamma_{1\%,1mm}$ [%][a] | Baseline | 99.4±0.8 [97.1 - 100] | 99.4±0.8 [96.9 - 100] | 99.4±0.7 [97.3 - 100] | 99.0±1.1 [95.4 - 99.9] |
| | Baseline +FLAIR | 99.5±0.7 [97.4 - 100] | 99.5±0.7 [97.2 - 100] | 99.4±0.7 [97.4 - 100] | 99.4±0.8 [97.2 - 100] |
| | Domain Randomisation | 99.4±0.8 [97.0 - 100] | 99.4±0.8 [96.9 - 100] | 99.3±0.8 [97.2 - 100] | 99.2±0.9 [96.6 - 99.9] |
| DD [%][b] | Baseline | -0.1±0.2 [-0.5 - 0.1] | -0.1±0.2 [-0.5 - 0.1] | -0.1±0.2 [-0.4 - 0.8] | 0.4±0.6 [-1.0 - 1.5] |
| | Baseline +FLAIR | -0.02±0.2 [-0.4 - 0.4] | -0.01±0.2 [-0.7 - 0.5] | 0.01±0.3 [-0.4 - 1.1] | 0.01±0.4 [-1.4 - 0.7] |
| | Domain Randomisation | -0.1±0.2 [0.5 - 0.2] | -0.2±0.2 [0.5 - 0.1] | -0.1±0.3 [-0.5 - 0.9] | 0.3±0.5 [-0.4 - 1.4] |

[a]Calculated in the D>10% prescribed dose. [b]Calculated in the D>90% prescribed dose.

The DD in treatment plans from FLAIR-based sCT was ≤1.5% and > 1% for three patients (PT2, PT13 and PT18). Specifically, a discrepancy between sCT and CT HU values was found for PT2 near the high-dose region: the skull near the frontal lobe was too thinly on sCT, causing HU values to be lower than in the CT. Discontinuities were visible in the skull of this post-surgical patient in the problematic area, although no part of the skull had been resected. For PT13, the high-dose region was located in the dorsal part of the cerebrum, where differences in skull thickness occurred between the FLAIR-based sCT generated by the Baseline model and the acquired CT, this time with higher HU values in the sCT than in the acquired CT. Notable dose differences were observed for PT18 near the nasal cavities, close to one of the isocentres of irradiation. The sCT generated by the Baseline model from this patient's FLAIR image revealed more prominent differences between HU values of sCT and acquired CT than the sCT generated for the other sequences.

Boxplots presenting the results for the DVH analysis are shown in Supplementary Material III.D for all three models. In general, minor differences in $D_{max}$ and $D_{median}$ were observed for OARs in DVH analysis for each sequence for the Baseline model. On average, differences were below 0.5% for every sequence and DVH point. Individually, most patients had differences in DVH points ≤1%. Exceptions for FLAIR-based sCT were the pituitary gland (PT14), optic chiasm (PT1), and lens (PT12), with differences ≤2%. PT12 patient had an RT plan with a vast irradiated area, matching the patient's large tumour volume. For this patient, for sCT derived from every MRI sequence, notable dose differences were observed around the body contour on the right half of the body.

*3.2.2. Baseline+FLAIR vs Baseline*

Training the Baseline+FLAIR model took 46.2 h with the application of early stopping at 450,000 iterations. The MAE obtained on T1w(Gd) (T1w: 65.5±7.7 HU; T1wGd: 64.6±9.3 HU), and T2w (71.2±8.8 HU) images was slightly worse for the Baseline+FLAIR model than the Baseline model (Fig. 3). For T2w and T1w, but not for T1wGd images this difference was statistically significant.

The most notable change in MAE was found on FLAIR images, favouring the Baseline+FLAIR model. Adding FLAIR images to the training data reduced the MAE from 106±20.7 HU to 72.6±10.1 HU (p< 0.5). Results for SSIM and PSNR were generally in line with the MAE.

For the Baseline+FLAIR model, $\gamma_{1\%1mm}$-pass rates were >97% for each patient and MRI sequence (Table 3). As for the Baseline model, pass rates $\gamma_{3\%3mm}$, and $\gamma_{2\%2mm}$ were all >99%. For FLAIR images, the Baseline+FLAIR model outperformed the Baseline model in terms of $\gamma_{1\%1mm}$-pass rate: 99.4±0.8% (Baseline+FLAIR model) vs 99.0±1.1% (Baseline model).

Likewise, the other two pass rates were significantly higher for the Baseline+FLAIR model. Surprisingly, despite the higher MAE obtained in T1w images for the Baseline+FLAIR model versus the Baseline model, a significantly higher $\gamma_{1\%1mm}$-pass rate was obtained for the Baseline+FLAIR model (99.5±0.7% vs 99.4±0.8%). All other differences in pass rates between the two models were not significant.

The absolute DD values obtained per sequence for the Baseline+FLAIR model were smaller than those obtained for the Baseline model (p<0.05), with DD<1.5% for every patient and seen sequence. For FLAIR images, the number of patients with a DD>1% was reduced to one (PT2) compared to three for the Baseline model. Similar to what was found for the Baseline model, for PT2, discrepancies between sCT and CT HU values were present near the high-dose region in the area of surgical intervention.

As for the Baseline model, differences in $D_{max}$ and $D_{median}$ were minor for all DVH points evaluated and all sequences, with average differences <0.5% and DVH point difference <1%, except for the cochlea of PT12 (≤1.5%) probably due to body contour mismatches on the right side.

*3.2.3. Domain Randomisation vs baselines*

The training time for the Domain Randomisation model was 66.4 h (450,000 iterations).

For the seen sequences, the MAE obtained for the Domain Randomisation model was higher (T1w: 67.6±7.4 HU; T1wGd: 66.5±9.2 HU; T2w: 71.5±7.9 HU) than that obtained for the Baseline and Baseline+FLAIR models (Fig. 3). All differences between the Domain Randomisation model and the Baseline model for these three sequences were statistically significant. Likewise, the differences between the Domain Randomisation and the Baseline+FLAIR model were statistically significant for T1w and T1wGd images but not for T2w images. Results for SSIM and PSNR were generally consistent with the MAE.

The MAE obtained for the Domain Randomisation model on FLAIR images (99.0±14.9 HU) was 7 HU lower than that obtained for the Baseline model (p<0.05), a difference which is larger than the increase in MAE obtained for the other sequences (T1w: +3 HU, T1wGd: +3 HU; T2w: +2 HU). Despite this decrease in MAE on FLAIR images obtained through the addition of RC images during network training, the MAE obtained for the Domain Randomisation model was 26 HU higher than that achievable when adding FLAIR images to the training dataset (Baseline+FLAIR model; p<0.05).

Figure 4 shows results for an example case generated by the Domain Randomisation model. Typical problematic areas for all sequences are the skull border and the nasal cavities. Similar difficult areas are observed for the Baseline and Baseline+FLAIR model. For FLAIR images specifically, the Domain Randomisation model produced sCTs on which the skull is thicker than on the acquired CT, showed as bright blue colour in the difference image (Fig. 4, right). Additionally, the musculature in the back of the neck is typically a problematic area for FLAIR images (arrow). Similar observations are made for the Baseline model. For the Baseline+FLAIR model, the overestimated skull thickness and neck musculature inaccuracies are less prominent.

Overall, a visual inspection of the results generated by the Domain Randomisation model from FLAIR images reveals that the model might be more robust than the Baseline model. Figure 5 shows three example patients for whom the Baseline model produced artefacts in the sCT (green rectangles). Such artefacts were not observed in the FLAIR-based sCTs produced by the Baseline+FLAIR and Domain Randomisation models. The bottom row shows images of a patient with an oedematous area in the frontal lobe. The area is hypointense on the FLAIR image, leading to an intensity similar to air in the sCT generated by the Baseline model, which translates to a high positive value in the image with the difference between CT and sCT. Results for the Domain Randomisation and Baseline+FLAIR model are less problematic.

Figure 5 also shows that the neck muscles are better depicted by the Domain Randomisation model than the baselines. Nevertheless, the smallest differences between CT and sCT in this area are observed in the sCT generated by the Baseline+FLAIR model. The lower differences between sCT and CT in muscle tissue compared to Baseline for the Domain Randomisation model was observed for all patients in the test set. A problem in FLAIR-based sCTs that remains unresolved after the application of domain randomisation is the mapping of the skull. Like the Baseline model, the Domain Randomisation model systematically produced sCTs with the skull mapped thicker than on the acquired CT, which is not observed for the Baseline+FLAIR model.

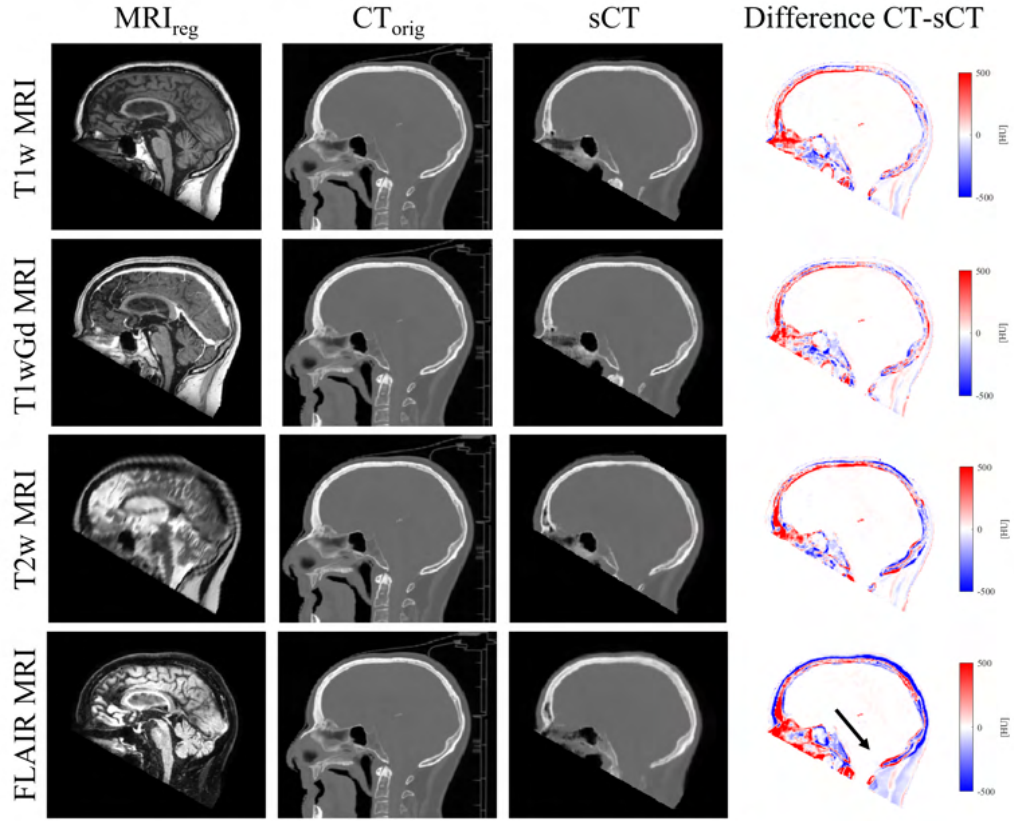

Figure 4: Results generated by the Domain Randomisation model, for a subject with average performance for T1w, T1wGd, T2w and FLAIR input images (top to bottom). The image shows from left to right: the original MRI, ground truth CT, sCT generated by the Domain Randomisation model, and the difference between the acquired CT and sCT. Typical problematic areas are the nasal cavities and the borders of the skull (bright in the image with the difference between CT and sCT on the right). For FLAIR specifically, the back of the neck (arrow) is problematic, and the skull is too thick on sCT, represented by the blue colour in the image with the difference between CT and sCT (right).

The $\gamma_{1\%1mm}$ for the Domain Randomisation model were >96% for each patient and each MRI sequence (Table 3). For the $\gamma_{3\%3mm}$, $\gamma_{2\%2mm}$, pass rates were all >99%, as for the other two models. Differences in $\gamma$-pass rates between the Baseline and Domain Randomisation model were insignificant for the seen sequences. However, for FLAIR images, the Domain Randomisation model outperformed the Baseline model for $\gamma_{1\%1mm}$: 99.2±0.9% vs 99.0±1.1% (p< 0.05).

Compared to the Baseline+FLAIR model for FLAIR images, the Domain Randomisation model resulted in significantly lower $\gamma_{1\%1mm}$ (99.2±0.9% vs 99.4±0.8%). Additionally, higher $\gamma_{1\%1mm}$ and $\gamma_{3\%3mm}$-pass rates were obtained for the Baseline+FLAIR model than for the Domain Randomisation model for T1w images (p< 0.05).

Differences in DD between the Domain Randomisation model and the Baseline model were not significant. Comparing the DD obtained for the Domain Randomisation and Baseline+FLAIR models resulted in p-values <0.05 for every sequence, with the DD obtained for the Base-

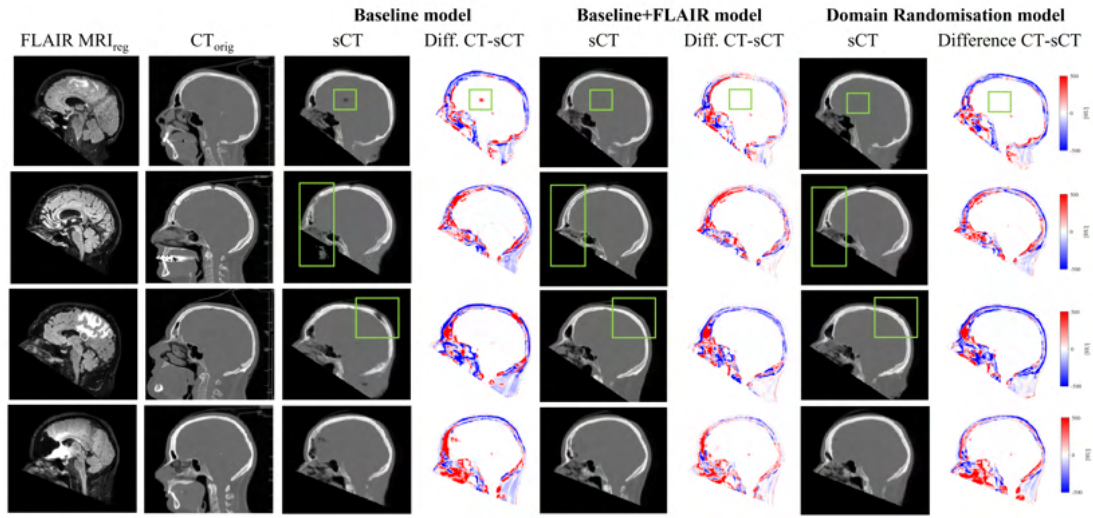

Figure 5: Results from FLAIR images generated by the three models for four different patients. Images from left to right: original FLAIR MRI, ground truth CT, sCT, and the difference between the acquired CT and sCT for the Baseline model, Baseline+FLAIR model and Domain Randomisation model, respectively. The areas marked with a green rectangle highlight artefacts in sCT produced by the Baseline model that are not present in the sCT generated by the Baseline+FLAIR and Domain Randomisation models. The bottom row shows an example patient with oedema in the frontal lobe. This area is hypointense on the FLAIR image, leading to problems in the sCT generated by the Baseline model.

line+FLAIR model smaller in absolute terms.

For the Domain Randomisation model, the DD in treatment plans from FLAIR-based sCT was < 1.5% and > 1% only for three patients (PT13, PT16 and PT18). For PT13 and PT18, we observed differences in the same regions already reported for the Baseline model. For PT16, dose differences in the high-dose region were substantial along the inner border of the skull, in line with the general observation that the MAE along the skull border was comparatively high for sCT generated from FLAIR images.

In general, boxplots for differences in DVH points for OARs reveal slight differences in $D_{max}$ and $D_{median}$ for all DVH points and sequences, with average differences < 0.5% as for the other two models. On an individual basis, most patients had differences in DVH points ≤ 1% for every OAR, except for coclhea of PT6 (<1.5%) and pituitary gland and lens of PT12 (<2.5%), which is comparable to the differences observed also for the baselines.

## 4. Discussion

In this work, we investigated the influence of domain randomisation for MRI-to-sCT generation, considering whether the generalisation to contrasts unseen during training can be increased.

We considered a cGAN model that after optimisation achieved image similarity to CT on par with models presented in the literature (Supplementary Material IV), where the reported range for T1w images is 45.4 HU±8.5 HU [57] to 131±14.3 HU [58] and 44.6±7.5 HU [57] to 89.3±10.3 HU [59] for T1wGd images. The accuracy of dose plans generated from sCT was high for all models and sequences. Considering $\gamma_{3\%3mm}$ and $\gamma_{2\%2mm}$-pass rates for clinical acceptability, dose plans were acceptable for all patients and sequences, even for FLAIR-based sCT generated by a model trained only on a mix of T1w(Gd) and T2w images (Baseline model). Altogether, the number of patients included in our training set (n=60) is significantly larger than the median number in other studies (n=33), we believe that our models' performance for the seen sequences is sufficient to explore generalisation, considering the following limitations.

A supervised framework was adopted, requiring a set of well-registered MRI-CT pairs. Poor registration between the CT and MRI in the training dataset is detrimental for models' performance [60]. We recurred to sole rigid registration, resulting in misregistration for the body contours, sinuses, and vertebrae for all the sequences. In the future, non-rigid registrations could be explored to improve the overall model performance or recurring to unpaired training [61].

Hyperparameter tuning was performed using only T1w images, which may not be optimal when mixing other sequences or applying domain randomisation. However, after performing a quick check of the hyperparameters on the RC model, we found that MAE could be decreased by 2-3 HU through optimisation, which was deemed minimal. Also, the adopted study design without hyperparameter optimization allows isolating the effect of domain randomisation.

A critical note should be made about our dose evaluation. Differences in the FOV of the acquired MRI and planning CT led to equal differences between the sCT and planning CT. Water-filling was used to avoid dose differences arising from different FOVs. However, this means that the dose accuracy of the sCT could be overestimated for beams passing through the water-filled area. In this sense, the sCT developed in this work should not be considered for clinical use but can be valuable to shed light on model generalisation.

Overall, we found the performance of a model trained on a mix of T1w(Gd) and T2w images (Baseline) was inferior on FLAIR images compared to performance on the other sequences. We found that domain randomisation leads to generating sCT from FLAIR images with significantly improved image similarity and dose accuracy compared to the Baseline model.

In an additional experiment [50], we found that the benefit of adding RC images to the training data was larger when only T1w acquired MRI were used for network training than a model trained on T1w(Gd) or T1w(Gd) and RC when both T1w(Gd) and T2w images were used (Domain Randomisation model vs Baseline model). Still, having at disposal the unseen sequence was not matched by the domain randomisation method. This means that whenever available, the MRI sequence that will be clinically used for MRI-only RT should be preferred. In case such a sequence is not available, domain randomisation can be considered as a method to increase model robustness. It seems that, at the moment, domain randomisation does not lead to strong contrast-agnostic method, which contrasts with what was claimed in the original work by Billot et al. [39]. We need to consider that compared to the original work (segmentation), we applied domain randomisation on a different and more challenging task (image synthesis). Also, Billot *et al.* [39] did not compare with a statistical test the performance on FLAIR to the performance on other sequences, which complicates judging to what extent their model is contrast-agnostic. Also, we speculate that our methods relied on generating RC images from labels, which may result in a loss of within-label structure that may be detrimental for the network considering the image synthesis task. Also, the adopted segmentations were not perfectly aligned with the corresponding ground truth (acquired CT), unlike in [38, 39], which might have reduced the effect of the RC images on network performance in our work. A solution may recur to obtaining through manual segmentation; we considered this procedure too expensive and out of scope for this research. Still, with the chosen study design, we were able to investigate the impact of domain randomisation. Future studies could clarify whether more accurate or elaborate label maps are more suitable as the basis for RC images, especially if the final goal of domain randomisation is obtaining a contrast-agnostic model.

A second domain randomisation method, based on linear combinations of acquired T1w(Gd) and T2w images, was proposed and tested for the first time. An advantage of this method over RC images is that it requires minimal effort and is easily applicable if multiple sequences are available per patient. However, the contrast produced by this method is less variable compared to RC, which could explain why this method is not as effective as using RC images. Theory and earlier studies suggest that variability beyond what the network will encounter in reality can be beneficial [41, 40, 62], in line with findings in [39], where synthetic images mimicking specific MRI sequences proved counterproductive. Future work could explore more elaborate domain randomisation methods, i.e., extending LC to non-linear combinations or increasing the number of acquired MRI sequences used for combination. Furthermore, the variability in the RC images could be further improved, e.g., using random elastic deformations or simulation of bias field artefacts as in [38, 39]. Another approach could explore using GANs or other DL models to generate synthetic training data, as suggested in, e.g., [63, 64].

A relevant question is whether the proposed domain randomisation approach could already be employed clinically to bridge smaller domain gaps than an entirely new sequence, like same-sequence data from a different hospital or changes in the acquisition protocol that might occur over time. Further evaluations on new datasets are needed to investigate whether this is the case.

This work provides the first attempt toward sCT generation from different MRI contrasts for MRI-only RT planning. A clear improvement was found in image similarity for sCT generated from an unseen sequence by Domain Randomisation models compared to baselines. Interestingly, in terms of dose accuracy, our baselines already achieved good results for most patients for the unseen sequence simply by training on a mix of other sequences. The Domain Randomisation model improved the $\gamma$-pass rate for the unseen sequence. In contrast, differences with the Baseline model in dose metrics were not statistically significant for the seen sequences, leading us to believe that the small decrease in image similarity obtained for the seen sequences is clinically acceptable. Moreover, the Domain Randomisation model reduced artefacts observed in FLAIR-based sCT comparable to the one observed from the Baseline model. Altogether, the results indicate that domain randomisation can improve generalisation to unseen sequences for sCT generation. Before clinically implementing the methods described in this work, dose accuracy must be evaluated in a clinical setting on MRI acquired with a larger FOV.

The results obtained in this work indicate that domain randomisation might help avoid the need for network re-training if the model is to be used on a sequence unseen during network training. This could be helpful if exceptions need to be made in imaging protocols for specific patients, e.g., due to possible allergic reactions to contrast agent or claustrophobia. On the other hand, each centre should determine whether the performance improvement found in this work is substantial enough to justify the effort associated with implementing domain randomisation, i.e. need of segmentations, in case alternative sequences are not already available.

## 5. Conclusion

We investigated the ability of a DL model to generate sCT on unseen sequences accurate for MRI-only radiotherapy. We considered two methods for domain randomisation, showing that adding random contrast images generated from label maps to the training data is more effective than applying random linear combinations of acquired MRI.

Generally, a satisfactory dose accuracy was obtained when training on a mix of acquired sequences, even for the unseen sequence. The adopted domain randomisation method improved dose accuracy and image similarity on this unseen sequence, but could not overperform having at disposal the unseen sequence during training. Domain randomisation can increase model robustness to unseen sequences, reducing the need for model re-training.

# Supplementary Material of 'Exploring contrast generalisation in deep learning-based brain MRI-to-CT synthesis'

Lotte Nijskens[a,b], Cornelis A.T. van den Berg[a,b], Joost J.C. Verhoeff[b], Matteo Maspero[a,b,*]

[a]*Computational Imaging Group for MR Diagnostics & Therapy, Center for Image Science, University Medical Center Utrecht, Heidelberglaan 100, Utrecht, 3584CX, The Netherlands*
[b]*Department of Radiotherapy, University Medical Center Utrecht, Heidelberglaan 100, Utrecht, 3584CX, The Netherlands*

**Abstract**

This supplementary material presents additional details regarding model optimisation (Sec. 1), the segmentation method adopted to create training set for the domain randomisation approach (Sec. 2), additional results regarding image similarity (PSNR, SSIM) and statistical analysis (Sec. 3), and a final section presenting an overview of the literature on sCT generation for brain patients and the methodology used to perform such search (Sec. 4).

## Contents

---

[*]Corresponding author
*Email address:* m.maspero@umcutrecht.nl (Matteo Maspero)
*URL:* https://orcid.org/0000-0003-0347-3375 (Matteo Maspero)

# 1. Model optimisation

## 1.1. Hyperparameter tuning

Details about the hyperparameter optimisation process are provided here. Hyperparameter tuning was done using a subset of 10 patients from the training dataset and the validation set (n = 10), using only T1w images without Gadolinium contrast. The optimisation was done by training 5,000 (coarse search) or 25,000 (refined search) iterations. Here, an iteration is defined as passing one image patch per patient through the network times the number of patients in one batch. The MAE between $CT_{crop}$ and generated sCT was leading in optimisation. SSIM and PSNR were used to decide if no differences in MAE were observed between models. Additionally, images were visually inspected for artefacts and image quality.

Hyperparameter optimisation was done through a grid search strategy. Hyperparameters considered were: (1) *initialisation method*, (2) *optimiser* and (3) corresponding *weight decay* value for the AdamW optimiser [1], (4) *patch size*, (5) the value of $\lambda$ in the loss function, (6) *learning rate*, (7) *batch size*, and (8) *number of downsampling steps* used in the U-Net generator architecture. Because patch-based inference was used, (9) the *blend mode* for patch combination and (10) the amount of *patch overlap* were also tuned. Table 1 contains the grid values considered.

Table 1: Hyperparameters considered for optimisation and the corresponding tested grid values.

| Hyperparameter | Tested grid values |
| --- | --- |
| Initialisation method | [Kaiming, Xavier] |
| Optimiser | [Adam, AdamW] |
| *Weight decay for AdamW* | [0.01, 0.1, 0.5] |
| Patch size | [32, 64, 128, 256] |
| $\lambda$ | [1, 10, 100, 500, 1,000, 5,000, 10,000] |
| Learning rate | [0.0001, 0.001, 0.01] |
| Batch size | [1, 5, 10] |
| Number of downsampling steps | [5, 6, 7] |
| Blend mode | [Gaussian, constant] |
| Patch overlap | [0.25, 0.5, 0.75] |

As a final optimisation step, three different decay strategies for the learning rate were compared, and early stopping was investigated. In this step, training was done on 30 training patients. The decay strategies considered were a constant learning rate of 0.001, a stepwise decaying learning rate and a cyclic learning rate. Two decay steps were applied for the stepwise decaying learning rate with decay factor $\gamma$ = 0.2, after 200,000 and 300,000 iterations. The cyclic learning rate was implemented with an initial constant phase at a learning rate of 200,000 iterations, followed by linear decay to 0 in 200,000 iterations and two cycles consisting of a restart to a learning rate of 0.01 with decay to 0 in 300,000 iterations. Early stopping was applied by selecting the first iteration for which the MAE in the intersection of the body contours did not improve for the subsequent three iterations considered. Note that evaluation was performed every 20,000 iterations.

## 1.2. Hyperparameter finetuning and dataset balancing

This section explains the final optimisation step. The ratio between T1w images with/without contrast and T2w images in the training set was balanced, and the batch size was finetuned for the mix of input sequences. The ratios T1w:T1wGd:T2w = 1:1:2 or T1w:T1wGd:T2w = 1:1:1 were compared by re-training the 3D model on a subset of fifteen patients to balance the ratio between T1w images with/without contrast and T2w images in the training set. Values of 1 and 5 were considered for the batch size. In this step, decisions were based on the MAE obtained for the validation set, using T1w, T1wGd and T2w images, i.e., the seen sequences[1].

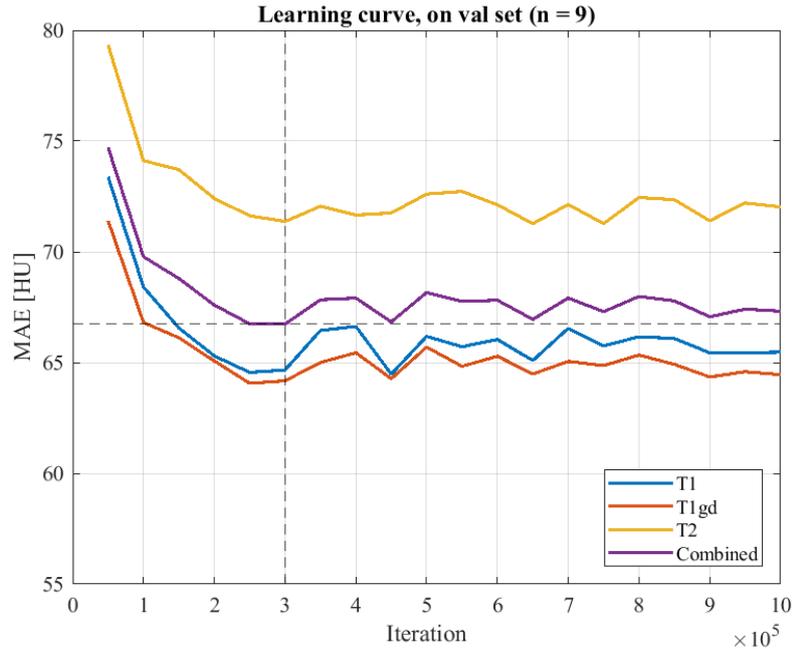

Figure 1: Learning curves for the Baseline model on the validation set (n = 9) for each sequence separately illustrating how early stopping was implemented in this work: the MAE in the intersection of the body contour at each evaluated iteration is plotted against the iteration number. The combined MAE is the average over T1w(Gd) and T2w images. The first iteration for which the combined MAE did not improve for the following three iterations was selected. The evaluation was done every 50,000 iterations.

A batch size of 1 was chosen. The ratio of T1w:T1wGd:T2w = 1:1:2 was adopted for the training dataset, meaning the whole training dataset (n = 60 patients) contained 60 T2w images, 30 T1w images and 30 T1wGd images.

*1.3. Example of the early stopping*

This section illustrates how early stopping was applied throughout this work, using the early stopping of the Baseline model as a reference (Fig. 1). The MAE was calculated for T1w, T1wGd and T2w images in the intersection of the body contours. Additionally, a combined MAE was calculated as the average for the three sequences. The first iteration for which this combined MAE did not improve for the following three iterations was selected, evaluating every 50,000 iterations. In this case, early stopping was applied at iteration 300,000 (dashed lines).

---

[1]One patient was retrospectively excluded from the validation set after failure of registration between the T2w image and the CT was observed. Validation of all models except the optimised 2D and 3D models was thus done on a nine-patient validation set.

For models trained using different acquired MRI (e.g., T1w(Gd) images only), the sequences taken into account when calculating the combined MAE were adjusted according to the training data.

## 2. Randomisation contrast method

### 2.1. Segmentation to label the training set

The methods for segmentation used to generate synthetic training data were broadly explained in the body. This section provides a more detailed explanation of the methods. Several methods were used to obtain an elaborate list of segmented structures (Table 2). Intracerebral structures were automatically segmented in T1w images using the open-source FastSurfer DL network [2]. The network requires input volumes of size [256, 256, 256]. Pre-processed T1w images were zero-padded if a dimension was less than 256 voxels, or zero-valued voxels were cropped from the volume in the case of larger dimensions to adhere to these sizes. After network inference, output label maps were reshaped to the original size of the T1w image. The segmentations of cerebrospinal fluid (CSF) and ventricles were grouped. OARs were added through automatic segmentation of T1w MRI using a previously in-house developed segmentation algorithm (unpublished) based on the DL model known as DeepMedic [3]. The model was previously developed for clinical use, employing the method as described in [4].

The GTV was added from an MRI-based clinical segmentation. Voxels in this GTV that had already been segmented as part of a cerebral structure with FastSurfer were assigned the corresponding FastSurfer label. Additionally, a clinical segmentation of the volume inside the skull was used to complement the segmentation of the CSF. All voxels falling inside the skull with no previous label (GTV, OAR, or from FastSurfer) were assigned the CSF label. A body contour was also obtained from the clinical segmentation. The intersection of this body contour and the registered MRI mask was used as a body mask for segmentation of some additional structures from $CT_{train}$.

Several structures were segmented using threshold operations on $CT_{train}$. Unless stated otherwise, thresholds were determined empirically based on one example patient from the training set and checked on two other patients. Background voxels and internal air were separated by applying a threshold of -0.5 to $CT_{train}$. The label for internal air was defined as those voxels falling inside the body mask but below the threshold. Segmentation of bone (i.e., bone and vertebrae) was achieved using a threshold of -0.120. Total bone was then subdivided into two classes (cortical bone and cancellous bone + bone marrow), using a threshold of 0.3216, with cortical bone defined as the voxels with an intensity above the threshold. The two bone labels were prioritised over the CSF label. Soft tissue was defined as all voxels inside the body mask that were not part of internal air and had not previously been assigned another label. Soft tissue was then divided into two classes (skin + muscle and other soft tissue) by thresholding the $CT_{train}$ (-0.2078). This threshold was determined using Matlab's implementation of Otsu's method [5]: automatic multi-threshold computation with four thresholds was applied to one example patient. The most appropriate threshold was then chosen by manually checking the result. Example of the label maps generated are visible in Figure 2.

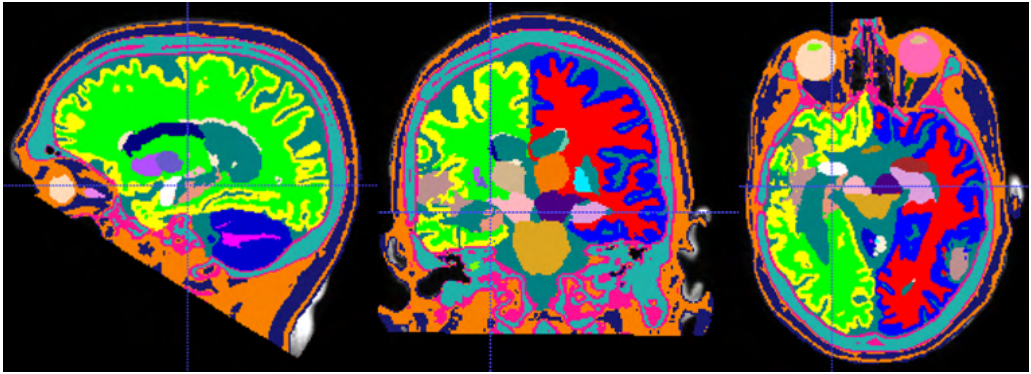

Figure 2: Example of a label map created through automatic segmentation of a single patient's T1w MRI and corresponding CT.

## 3. Additional results

*3.1. Random contrast vs linear combination*

*3.1.1. PSNR and SSIM*

As a complement to the results for MAE presented in the main body of this work, Table 3 presents the results for the SSIM. Table 4 presents the results for the PSNR.

*3.1.2. Statistical comparisons*

This section provides the p-values obtained with Wilcoxon-signed rank tests conducted for comparisons presented in the body of this work, both for image similarity metrics and for dosimetric accuracy for the Baseline, Baseline+FLAIR and Domain Randomisation models. Image similarity metrics obtained on the test set (n = 25) per model were compared for pairs of sequences (Table 5).

Also, image similarity metrics per sequence were compared for pairs of models between Baseline, Baseline+FLAIR and Domain Randomisation model (Table 6). For each of the three models (Baseline, Baseline+FLAIR and Domain Randomisation), dosimetric accuracy was statistically compared between sequences (Table 7). Additionally, comparisons were made in dosimetric accuracy between pairs of models, with p-values reported per sequence (Table 8).

Table 2: Look up table for the label maps and corresponding automatic segmentation method.

| Label | Structure | Segmentation method |
|---|---|---|
| 0 | Background | Outside body mask (clinical segmentation, MRI mask) |
| 1 | Cortical white matter (L) | FastSurfer [2] label 2 |
| 2 | Cortical white matter (R) | FastSurfer label 41 |
| 3 | Cortical grey matter (L) | FastSurfer labels 1000-1999 |
| 4 | Cortical grey matter (R) | FastSurfer labels $\geq$ 2000 |
| 5 | Cerebellar white matter (L) | FastSurfer label 7 |
| 6 | Cerebellar white matter (R) | FastSurfer label 46 |
| 7 | Cerebellar cortex (L) | FastSurfer label 8 |
| 8 | Cerebellar cortex (R) | FastSurfer label 47 |
| 9 | Thalamus (L) | FastSurfer label 10 |
| 10 | Thalamus (R) | FastSurfer label 49 |
| 11 | Caudate nucleus (L) | FastSurfer label 11 |
| 12 | Caudate nucleus (R) | FastSurfer label 50 |
| 13 | Putamen (L) | FastSurfer label 12 |
| 14 | Putamen (R) | FastSurfer label 51 |
| 15 | Pallidum (L) | FastSurfer label 13 |
| 16 | Pallidum (R) | FastSurfer label 52 |
| 17 | Hippocampus (L) | FastSurfer label 17 |
| 18 | Hippocampus (R) | FastSurfer label 53 |
| 19 | Amygdala (L) | FastSurfer label 18 |
| 20 | Amygdala (R) | FastSurfer label 54 |
| 21 | Accumbens (L) | FastSurfer label 26 |
| 22 | Accumbens (R) | FastSurfer label 58 |
| 23 | Ventral diencephalon (L) | FastSurfer label 28 |
| 24 | Ventral diencephalon (R) | FastSurfer label 60 |
| 25 | Choroid plexus (L) | FastSurfer label 31 |
| 26 | Choroid plexus (R) | FastSurfer label 63 |
| 27 | White matter hypointensities (if present) | FastSurfer label 77 |
| 28 | Brain stem | FastSurfer label 16 |
| 29 | CSF | FS 4, 5, 14, 15, 24, 43, 44 + voxels inside skull (clinical segmentation) without other label |
| 30 | GTV | Clinical segmentation, not labelled by FastSurfer |
| 31 | Eye (L) | DeepMedic-based OAR segmentation network |
| 32 | Eye (R) | DeepMedic-based OAR segmentation network |
| 33 | Lense (L) | DeepMedic-based OAR segmentation network |
| 34 | Lense (R) | DeepMedic-based OAR segmentation network |
| 35 | Cochlea (L) | DeepMedic-based OAR segmentation network |
| 36 | Cochlea (R) | DeepMedic-based OAR segmentation network |
| 37 | Lacrimal gland (L) | DeepMedic-based OAR segmentation network |
| 38 | Lacrimal gland (R) | DeepMedic-based OAR segmentation network |
| 39 | Optic nerve (L) | DeepMedic-based OAR segmentation network |
| 40 | Optic nerve (R) | DeepMedic-based OAR segmentation network |
| 41 | Pituitary gland | DeepMedic-based OAR segmentation network |
| 42 | Optic chiasm | DeepMedic-based OAR segmentation network |
| 43 | Bone: Cortical bone | Thresholding $CT_{train}$ |
| 44 | Bone: cancellous bone + bone marrow | Thresholding $CT_{train}$ |
| 45 | Soft tissue: muscle + skin | Thresholding $CT_{train}$ |
| 46 | Soft tissue: other soft tissue | Thresholding $CT_{train}$ |
| 47 | Internal air | Thresholding $CT_{train}$, inside body mask |

Table 3: SSIM for sCT generated by the RC and LC models per MRI sequence. Metrics were calculated on the validation set (n = 9) within the intersection of the body contour of the sCT and CT. Mean values and standard deviations ($\mu \pm 1\sigma$) and range ([min - max]) are reported. Wilcoxon-signed rank tests were used for statistical comparisons. Values of p < 0.05 were regarded as statistically significant.

| Metric | Model | Sequence | | | |
|---|---|---|---|---|---|
| | | T1w | T1wGd | T2w | FLAIR |
| SSIM | RC | 0.869 ± 0.027 [0.801 - 0.889] | 0.873 ± 0.027 [0.804 - 0.895] | 0.855 ± 0.024 [0.804 - 0.887] | 0.803 ± 0.047 [0.720 - 0.865] |
| | LC | 0.870 ± 0.026 [0.806 - 0.893] | 0.871 ± 0.026 [0.805 - 0.891] | 0.854 ± 0.025 [0.799 - 0.880] | 0.793 ± 0.047 [0.709 - 0.864] |
| p-value | | 0.9 | 0.4 | 0.4 | 0.07 |

Table 4: PSNR for sCT generated by the RC and LC models per MRI sequence. Metrics were calculated on the validation set (n = 9) within the intersection of the body contour of the sCT and CT. Mean values and standard deviations ($\mu \pm 1\sigma$) and range ([min - max]) are reported. Wilcoxon-signed rank tests were used for statistical comparisons. Values of p < 0.05 were regarded as statistically significant.

| Metric | Model | Sequence | | | |
|---|---|---|---|---|---|
| | | T1w | T1wGd | T2w | FLAIR |
| PSNR [dB] | RC | 28.1 ± 1.3 [25.7 - 29.9] | 28.3 ± 1.4 [25.6 - 30.4] | 27.5 ± 1.2 [25.8 - 29.4] | 25.6 ± 1.5 [23.7 - 28.5] |
| | LC | 28.1 ± 1.3 [25.8 - 30.3] | 28.2 ± 1.4 [25.6 - 30.2] | 27.4 ± 1.1 [25.7 - 29.2] | 25.3 ± 1.7 [23.0 - 28.7] |
| p-value | | 1 | 0.6 | 0.4 | 0.07 |

Table 5: Statistical comparisons of image similarity metrics (MAE, PSNR, SSIM) between sequences for Baseline, Baseline+FLAIR and Domain Randomisation model. Results as p-values per model. Metrics were calculated on the test set (n = 25) within the intersection of the body contour of the sCT and CT. P-values were calculated with Wilcoxon-signed rank tests. Values of p < 0.05 were regarded as statistically significant.

| Metric | Comparison | Model | | |
|---|---|---|---|---|
| | | Baseline | Baseline+FLAIR | Domain Randomisation |
| MAE | T1 vs T1gd | 0.4 | 0.07 | 0.09 |
| | T1 vs T2 | 0.002 | $9*10^{-4}$ | 0.007 |
| | T1 vs FLAIR | $1*10^{-5}$ | $4*10^{-4}$ | $1*10^{-5}$ |
| | T1gd vs T2 | 0.001 | $5*10^{-4}$ | 0.001 |
| | T1gd vs FLAIR | $1*10^{-5}$ | 0.001 | $1*10^{-5}$ |
| | T2 vs FLAIR | $1*10^{-5}$ | 0.4 | $1*10^{-5}$ |
| SSIM | T1 vs T1gd | 0.2 | 0.09 | 0.1 |
| | T1 vs T2 | 0.003 | 0.001 | 0.004 |
| | T1 vs FLAIR | $1*10^{-5}$ | 0.004 | $1*10^{-5}$ |
| | T1gd vs T2 | $9*10^{-4}$ | $4*10^{-4}$ | $8*10^{-4}$ |
| | T1gd vs FLAIR | $1*10^{-5}$ | 0.003 | $1*10^{-5}$ |
| | T2 vs FLAIR | $1*10^{-5}$ | 0.5 | $1*10^{-5}$ |
| PSNR | T1 vs T1gd | 0.3 | 0.02 | 0.02 |
| | T1 vs T2 | $5*10^{-4}$ | $3*10^{-4}$ | $1*10^{-3}$ |
| | T1 vs FLAIR | $1*10^{-5}$ | 0.001 | $1*10^{-5}$ |
| | T1gd vs T2 | $3*10^{-4}$ | $1*10^{-4}$ | $2*10^{-4}$ |
| | T1gd vs FLAIR | $1*10^{-5}$ | 0.001 | $1*10^{-5}$ |
| | T2 vs FLAIR | $1*10^{-5}$ | 0.7 | $1*10^{-5}$ |

Table 6: Statistical comparisons of image similarity metrics between models for Baseline, Baseline+FLAIR and Domain Randomisation model. Results as p-values per sequence. Metrics were calculated on the test set (n = 25) within the intersection of the body contour of the sCT and CT. P-values were calculated with Wilcoxon-signed rank tests. Values of p < 0.05 were regarded as statistically significant.

| Metric | Comparison | Sequence | | | |
|---|---|---|---|---|---|
| | | T1w | T1wGd | T2w | FLAIR |
| MAE | Baseline vs Baseline+FLAIR | 0.007 | 0.1 | $3*10^{-4}$ | $1*10^{-5}$ |
| | Baseline vs Domain Randomisation | $2*10^{-4}$ | $3*10^{-5}$ | $2*10^{-4}$ | $3*10^{-5}$ |
| | Baseline+FLAIR vs Domain Randomisation | $4*10^{-4}$ | $5*10^{-4}$ | 0.4 | $1*10^{-5}$ |
| SSIM | Baseline vs Baseline+FLAIR | 0.054 | 0.3 | $5*10^{-4}$ | $1*10^{-5}$ |
| | Baseline vs Domain Randomisation | 0.002 | $2*10^{-4}$ | 0.002 | $3*10^{-4}$ |
| | Baseline+FLAIR vs Domain Randomisation | 0.003 | $9*10^{-4}$ | 1 | $1*10^{-5}$ |
| PSNR | Baseline vs Baseline+FLAIR | 0.007 | 0.07 | $4*10^{-4}$ | $1*10^{-5}$ |
| | Baseline vs Domain Randomisation | $7*10^{-4}$ | $2*10^{-4}$ | 0.001 | 0.002 |
| | Baseline+FLAIR vs Domain Randomisation | 0.002 | 0.007 | 0.6 | $1*10^{-5}$ |

Table 7: Statistical comparisons of metrics for dosimetric accuracy between sequences for Baseline, Baseline+FLAIR and Domain Randomisation model. Results as p-values per model. Metrics were calculated on the test set (n = 25) within the intersection of the body contour of the sCT and CT. P-values were calculated with Wilcoxon-signed rank tests. Values of p < 0.05 were regarded as statistically significant.

| Metric | Comparison | Model | | |
| --- | --- | --- | --- | --- |
| | | Baseline | Baseline +FLAIR | Domain Randomisation |
| $\gamma_{3\%,3mm}$ | T1 vs T1gd | 0.6 | 0.9 | 0.4 |
| | T1 vs T2 | 0.8 | 1 | 1 |
| | T1 vs FLAIR | 0.02 | 0.1 | 0.4 |
| | T1gd vs T2 | 0.6 | 1 | 0.7 |
| | T1gd vs FLAIR | 0.01 | 0.3 | 0.1 |
| | T2 vs FLAIR | 0.1 | 0.08 | 0.8 |
| $\gamma_{2\%,2mm}$ | T1 vs T1gd | 0.2 | 0.4 | 1 |
| | T1 vs T2 | 0.04 | 0.06 | 0.2 |
| | T1 vs FLAIR | 0.001 | 0.2 | 0.005 |
| | T1gd vs T2 | 0.3 | 0.3 | 0.1 |
| | T1gd vs FLAIR | 0.005 | 0.2 | 0.01 |
| | T2 vs FLAIR | 0.1 | 0.5 | 0.2 |
| $\gamma_{1\%,1mm}$ | T1 vs T1gd | 0.5 | 0.8 | 0.3 |
| | T1 vs T2 | 0.7 | 0.04 | 0.3 |
| | T1 vs FLAIR | $5*10^{-5}$ | 0.03 | $4*10^{-4}$ |
| | T1gd vs T2 | 0.4 | 0.1 | 0.09 |
| | T1gd vs FLAIR | $1*10^{-5}$ | 0.001 | $1*10^{-4}$ |
| | T2 vs FLAIR | $1*10^{-4}$ | 0.9 | 0.01 |
| DD | T1 vs T1gd | 0.2 | 0.4 | 0.2 |
| | T1 vs T2 | 0.4 | 0.9 | 0.2 |
| | T1 vs FLAIR | $2*10^{-4}$ | 0.2 | $2*10^{-5}$ |
| | T1gd vs T2 | 0.8 | 0.5 | 0.2 |
| | T1gd vs FLAIR | $2*10^{-4}$ | 0.8 | $3*10^{-5}$ |
| | T2 vs FLAIR | $4*10^{-4}$ | 0.4 | $1*10^{-5}$ |

Table 8: Statistical comparisons of metrics for dosimetric accuracy between models for Baseline, Baseline+FLAIR and Domain Randomisation model. Results as p-values per sequence. Metrics were calculated on the test set (n = 25) within the intersection of the body contour of the sCT and CT. P-values were calculated with Wilcoxon-signed rank tests. Values of p < 0.05 were regarded as statistically significant.

| Metric | Comparison | Sequence | | | |
|---|---|---|---|---|---|
| | | T1w | T1wGd | T2w | FLAIR |
| $\gamma_{3\%,3mm}$ | Baseline vs Baseline+FLAIR | 0.7 | 1 | 0.9 | 0.003 |
| | Baseline vs Domain Randomisation | 0.1 | 0.7 | 0.7 | 0.07 |
| | Baseline+FLAIR vs Domain Randomisation | 0.04 | 0.4 | 0.8 | 0.006 |
| $\gamma_{2\%,2mm}$ | Baseline vs Baseline+FLAIR | 0.8 | 0.2 | 0.5 | 0.01 |
| | Baseline vs Domain Randomisation | 0.4 | 0.7 | 0.6 | 0.3 |
| | Baseline+FLAIR vs Domain Randomisation | 0.7 | 0.4 | 0.9 | 0.053 |
| $\gamma_{1\%,1mm}$ | Baseline vs Baseline+FLAIR | 0.003 | 0.1 | 0.2 | $2*10^{-4}$ |
| | Baseline vs Domain Randomisation | 0.7 | 0.5 | 0.5 | 0.005 |
| | Baseline+FLAIR vs Domain Randomisation | 0.04 | 0.07 | 0.07 | $9*10^{-5}$ |
| DD | Baseline vs Baseline+FLAIR | 0.01 | $7*10^{-4}$ | 0.002 | $4*10^{-5}$ |
| | Baseline vs Domain Randomisation | 0.3 | 0.1 | 0.06 | 0.06 |
| | Baseline+FLAIR vs Domain Randomisation | 0.003 | $2*10^{-4}$ | $4*10^{-5}$ | $1*10^{-4}$ |

## 3.2. Baselines and randomisation
### 3.2.1. PSNR and SSIM

As a complement to the violin plot presenting results for MAE presented in the body (Fig. 3), Figs. 3 and 4 present results for PSNR and SSIM.

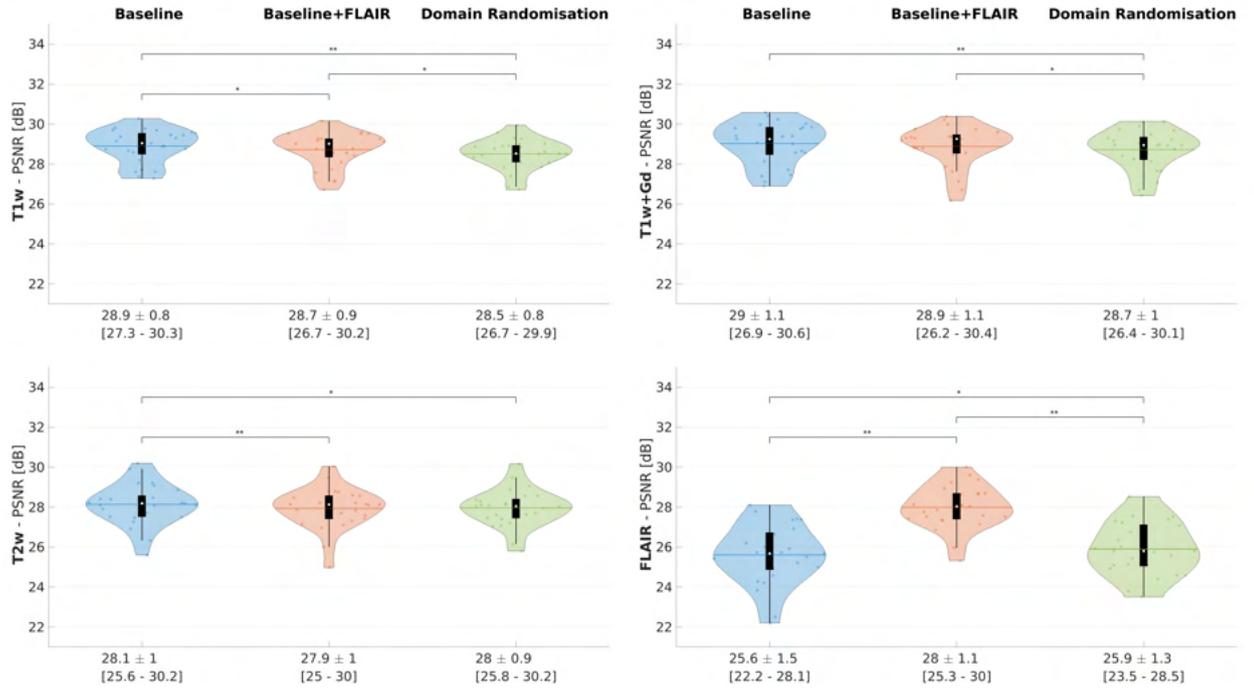

Figure 3: Violin and box-and-whisker plots for the PSNR in the intersection of the body contour of sCT compared to ground truth CT on the test set (n = 25) for sCT generated by the Baseline (blue), Baseline+FLAIR (orange) and Domain Randomisation model (green). Results are presented per sequence: T1w (top left), T1wGd (top right), T2w (bottom left) and FLAIR (bottom right). The black box indicates the interquartile range and median (white circle) with whiskers indicating the range, outliers excluded. The width of the violin indicates the distribution of the data points. The mean values and standard deviations are shown. Statistically significant differences are indicated by * ($p < 0.05$) or ** ($p < 0.001$). Wilcoxon-signed rank tests were used for statistical testing.

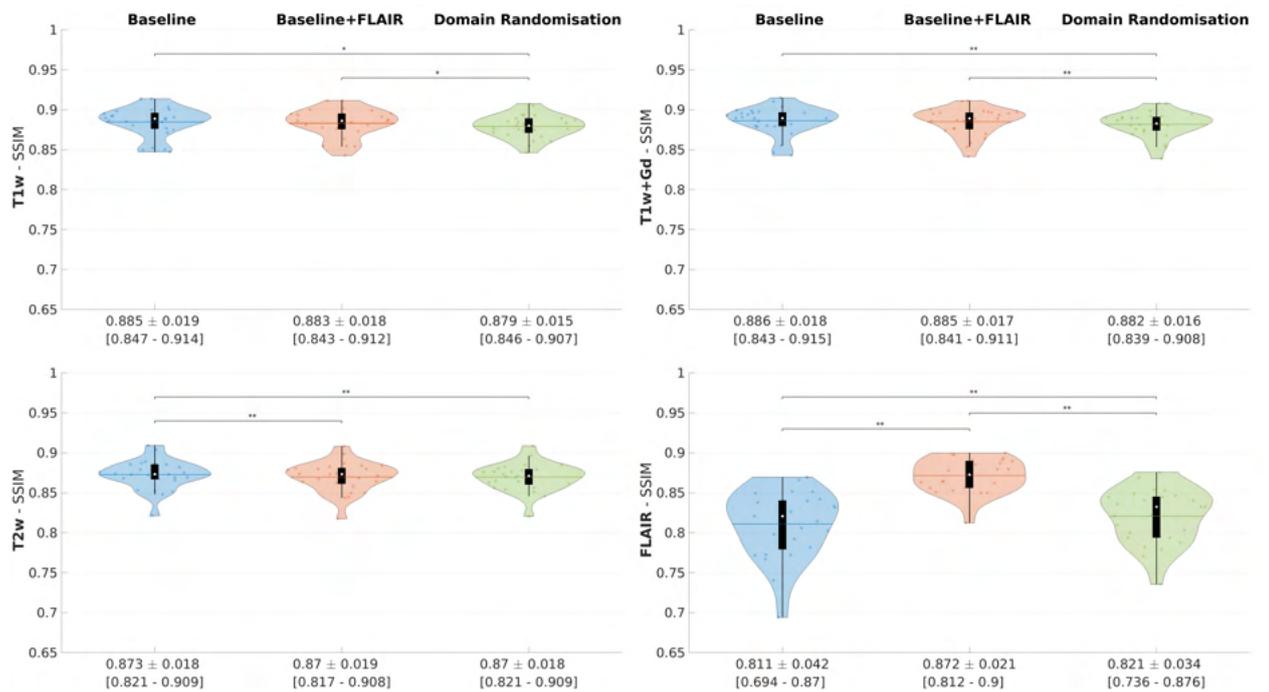

Figure 4: Violin and box-and-whisker plots for the SSIM in the intersection of the body contour of sCT compared to ground truth CT on the test set (n = 25) for sCT generated by the Baseline (blue), Baseline+FLAIR (orange) and Domain Randomisation model (green). Results are presented per sequence: T1w (top left), T1wGd (top right), T2w (bottom left) and FLAIR (bottom right). The black box indicates the interquartile range and median (white circle) with whiskers indicating the range, outliers excluded. The width of the violin indicates the distribution of the data points. The mean values and standard deviations are shown. Statistically significant differences are indicated by * ($p < 0.05$) or ** ($p < 0.001$). Wilcoxon-signed rank tests were used for statistical testing.

*3.2.2. Dosimetric accuracy for baselines and randomisation*

As a complement to the results for dose accuracy (3D $\gamma$-pass rate with 1%,1mm criterion and DD in the high dose region) presented in the body, Table 9 presents results for the $\gamma$-pass rates with 3%,3mm and 2%,2mm criteria.

Table 9: Dose evaluation ($\gamma_{3\%,3mm}$ and $\gamma_{2\%,2mm}$) for sCT generated by the Baseline, Baseline+FLAIR and Domain Randomisation models per MRI sequence. Dosimetric accuracy was assessed through plan re-calculation on water-filled sCT compared to the water-filled acquired CT. Mean values and standard deviations ($\mu \pm 1\sigma$) and range ([min - max]) are reported.

| Metric | Model | Sequence | | | |
|---|---|---|---|---|---|
| | | T1w | T1wGd | T2w | FLAIR |
| $\gamma_{3\%,3mm}$ [%][a] | Baseline | 99.99 ± 0.01 [99.96 - 100] | 100 ± 0.01 [99.9 - 100] | 99.99 ± 0.01 [99.97 - 100] | 99.99 ± 0.02 [99.9 - 100] |
| | Baseline +FLAIR | 100 ± 0.01 [99.95 - 100] | 100 ± 0.01 [99.97 - 100] | 99.99 ± 0.01 [99.96 - 100] | 100 ± 0.01 [99.97 - 100] |
| | Domain Randomisation | 99.99 ± 0.01 [99.95 - 100] | 99.99 ± 0.01 [99.95 - 100] | 99.99 ± 0.01 [99.96 - 100] | 99.99 ± 0.01 [99.95 - 100] |
| $\gamma_{2\%,2mm}$ [%][a] | Baseline | 99.95 ± 0.1 [99.7 - 100] | 99.95 ± 0.1 [99.6 - 100] | 99.9 ± 0.1 [99.7 - 100] | 99.9 ± 0.2 [99.4 - 100] |
| | Baseline +FLAIR | 99.96 ± 0.1 [99.7 - 100] | 99.95 ± 0.1 [99.7 - 100] | 99.9 ± 0.1 [99.7 - 100] | 99.95 ± 0.1 [99.7 - 100] |
| | Domain Randomisation | 99.95 ± 0.08 [99.7 - 100] | 99.95 ± 0.1 [99.6 - 100] | 99.9 ± 0.08 [99.7 - 100] | 99.9 ± 0.1 [99.5 - 100] |

[a]Calculated in the D > 10 % prescribed region. [b]Calculated in the D > 90 % prescribed region.

As part of the dosimetric evaluation of the generated sCT, DVH differences in $D_{median}$ and $D_{max}$ between sCT- and CT-based dose plans were evaluated for the brainstem, optic chiasm, lenses, cochleae and pituitary gland. Boxplots representing these differences are shown per sequence in Fig. 5 (Baseline model), Fig. 6 (Baseline+FLAIR model) and Fig. 7 (Domain Randomisation model). The main differences are discussed in the body.

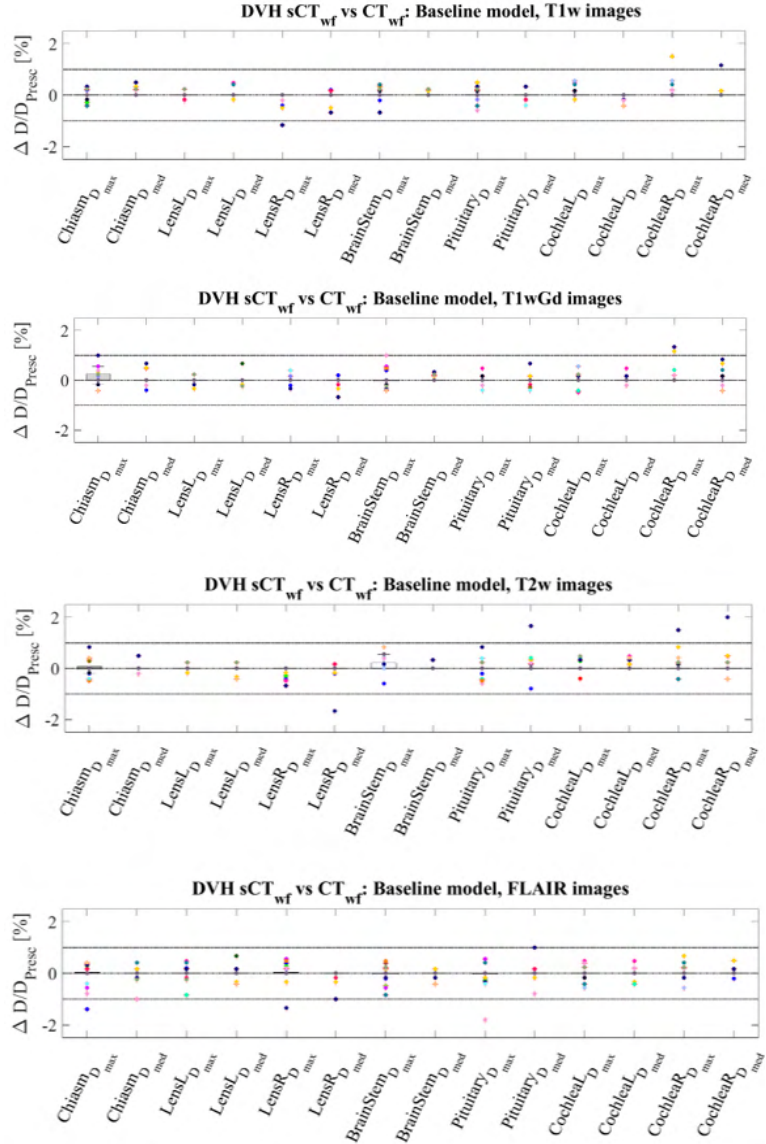

Figure 5: Boxplots for DVH differences between sCT$_{wf}$ generated by the Baseline model and CT$_{wf}$-based dose plans in D$_{max}$ and D$_{median}$ for OARs: brainstem, optic chiasm, lenses, cochleae and pituitary gland. Results are shown per sequence (top to bottom: T1w, T1wGd, T2w and FLAIR images). Dots represent outliers, with each colour representing a different patient.

## 4. Literature overview: brain sCT

*4.1. Search strategy*

On April 29th, 2022, a systematic literature search was done in the scientific database Scopus for studies about the accuracy of DL-based sCT generation from brain MRI for MR-only RT planning. The search strategy included the following keywords: (sCT OR "synth* CT" OR "CT synth*" OR pseudoCT OR "pseudo CT" OR "pseudo-CT" OR pCT OR "MRI-to-CT") AND (brain OR cerebr* OR head OR skull) and (GAN OR "generative adversarial net*" OR "deep learning" OR CNN OR Unet OR U-Net OR "neural net*"). The search was performed on title, abstract and keywords with no limitation on the publication date or language. The titles and abstracts were reviewed to select studies for full-text review. Any doubts about inclusion were resolved by screening the full text.

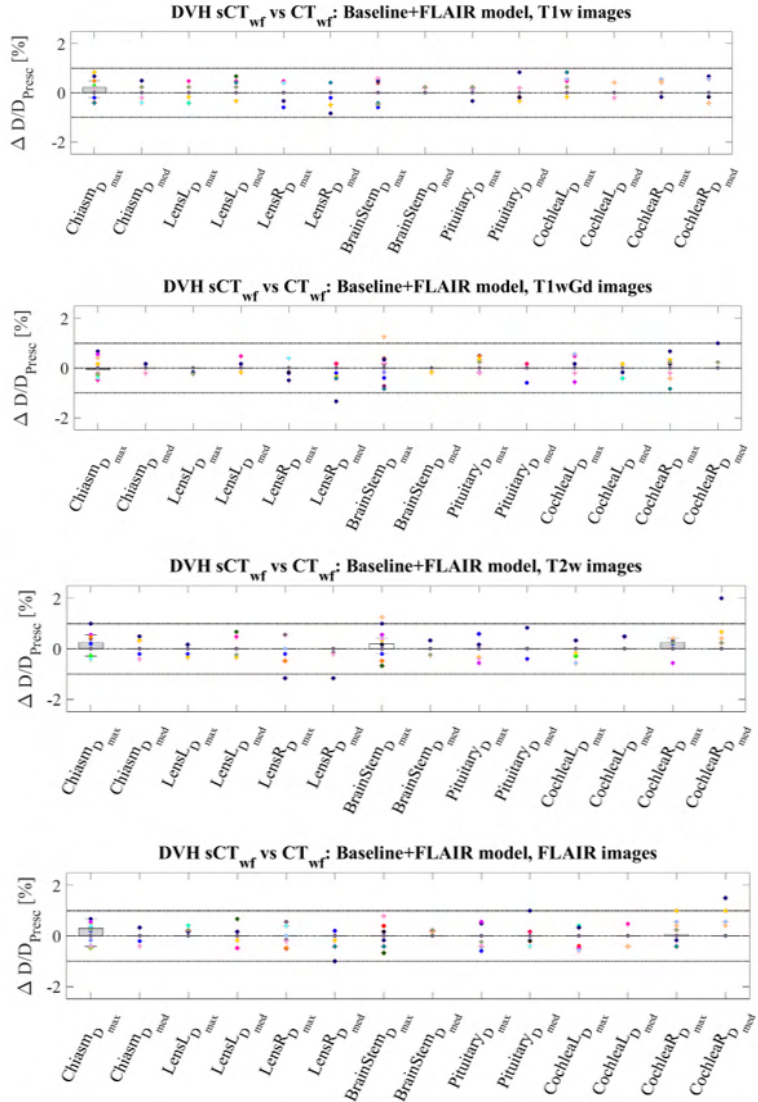

Figure 6: Boxplots for DVH differences between sCT$_{wf}$ generated by the Baseline+FLAIR model and CT$_{wf}$-based dose plans in D$_{max}$ and D$_{median}$ for OARs: brainstem, optic chiasm, lenses, cochleae and pituitary gland. Results are shown per sequence (top to bottom: T1w, T1wGd, T2w and FLAIR images). Dots represent outliers, with each colour representing a different patient.

The MAE obtained from the comparison of sCT to acquired CT had to be reported, possibly with additional metrics SSIM, PSNR and $\gamma$-pass rates with 3%,3mm, 2%,2mm or 1%1,mm criterion. Only studies aiming to generate sCT from brain MRI scans for MR-only RT planning were considered for inclusion. Articles were excluded if: a. the full text was unavailable in Dutch or English; b. the article was a conference paper or review; c. the aim was CT-to-MRI translation instead of MRI-to-CT translation; d. patients were solely head-and-neck cancer patients instead of brain cancer patients; e. either the input sequence was not a T1w, T1w+Gd, T2w or T2w FLAIR image, or the input sequence was undefined; or f. the article was a duplicate evaluation of a DL model already evaluated in an earlier study.

*4.2. Data extraction*

A spreadsheet was designed for data extraction, extracting the following information from the included articles: a. basic information, including the first author to allow identification, year of publication, journal; and b. data needed for comparison with results obtained in the current work: input MRI sequence, model configuration, type of model, image similarity metrics (MAE, SSIM, PSNR), and $\gamma$-pass rates for 3%,3mm, 2%,2mm and 1%,1mm criteria. Dose differences were not considered because of the large variability in reported metrics.

*4.3. Results*

Only three studies were identified that presented results for T2w FLAIR images separately. Most included studies (n = 17; Table 10) used T1w images as input sequence, followed by T1wGd images (n = 7) and T2w images (n = 5). The MAE obtained for models taking T1w images as input sequences ranged between 45.4 HU [6] to 131 HU [7], in addition to one exceptionally low MAE of 9.02 HU [8] (Table 10). For T1wGd images, mean MAEs ranged from 44.6 HU [6] to 89.3 HU [9]. Values between 45.7 HU [6] and 68.3 HU [10] were identified for T2w images. For T2w FLAIR images, two identified studies reported MAE values of 51.2 HU [6] and 59.3 HU [11], and one study reported an MAE of 115 ± 22. Additionally, in [12], a mean MAE of 61.9 ± 22.6 was obtained for a model trained on a mix of T2w images with and without FLAIR. However, no statistical comparisons were provided between groups of patients with different imaging protocols [12].

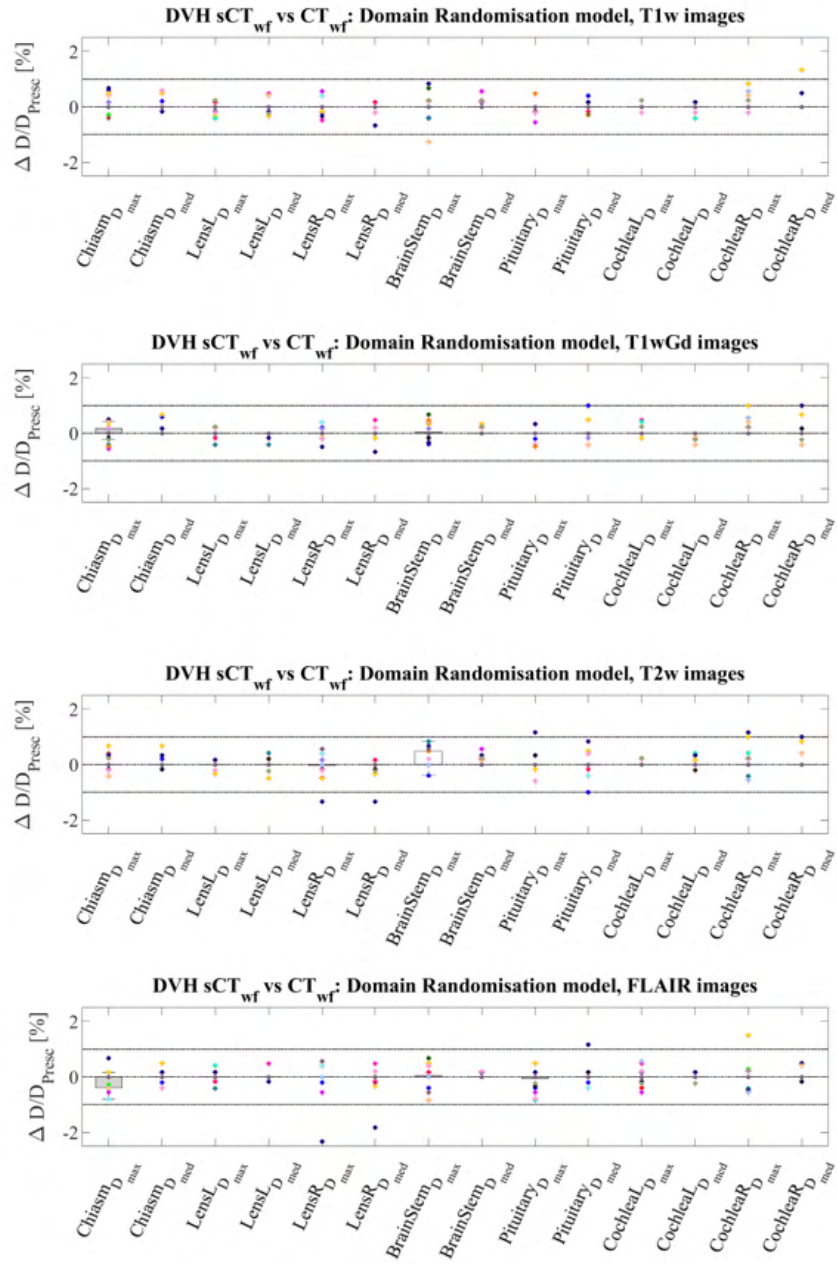

Figure 7: Boxplots for DVH differences between sCT$_{wf}$ generated by the Domain Randomisation model and CT$_{wf}$-based dose plans in D$_{max}$ and D$_{median}$ for OARs: brainstem, optic chiasm, lenses, cochleae and pituitary gland. Results are shown per sequence (top to bottom: T1w, T1wGd, T2w and FLAIR images). Dots represent outliers, with each colour representing a different patient.

Table 10: Literature review: brain sCT with image similarity metrics and gamma analysis.

| Training sequence | Reference | Pts[a] | Conf.[b] | Model | Image similarity | | | Gamma analysis | | |
|---|---|---|---|---|---|---|---|---|---|---|
| | | | | | MAE [HU] | SSIM | PSNR [dB] | $\gamma_{3\%,3mm}$ [%] | $\gamma_{2\%,2mm}$ [%] | $\gamma_{1\%,1mm}$ [%] |
| T1w | Han 2017 [13] | 18 | 2D | U-net | 84.8±17.3 | - | - | - | - | - |
| | Dinkla 2018 [14] | 26 | 2D+[c] | CNN | 67±11 | - | - | 99.9±0.2 | 99.1±0.80 | 97.0±2.2 |
| | Xiang 2018 [15] | 16 | 2.5D[d] | U-net | 85.4±9.24 | - | 27.3±1.1 | - | - | - |
| | Gupta 2019 [16] | 47 | 2D | U-net | 81.0±14.6 | - | - | - | - | - |
| | Koike 2019 [17] | 15 | 2Dp | GAN | 120±20.4 | - | - | 99.7±0.5 | 98.7±1.2 | 94.2±4.9 |
| | Lei 2019 [18] | 24 | 3Dp | GAN | 55.7±9.4 | - | 25.8±1.81 | - | - | - |
| | Neppl 2019 [19] | 57 | 2D | U-net | 116±26 | - | - | - | 98±2 | - |
| | Shafai-Erfani 2019 [20] | 25 | 3Dp | GAN | 54.6±6.81 | - | - | 99.96±0.21 | 98.4±3.51 | 90.8±7.8 |
| | Spadea 2019 [21] | 12 | 2D+[e] | U-net | 54±7 | - | - | - | - | - |
| | Alvarez-Andres 2020 [22][f] | 134 | 3Dp | CNN | 84±25 | - | - | 99.8±0.18 | 99.6±0.33 | 97.9±1.16 |
| | Massa 2020 [6] | 81 | 2D | U-net | 45.4±8.52 | 0.65±0.05 | 43.0±2.02 | - | - | - |
| | Xu 2020 [8] | 33 | 2D | GAN | 9.02±0.82 | 0.75±0.77 | - | - | - | - |
| | Irmak 2021 [7] | 20 | 2D | GAN | 131±14.3 | - | - | - | 99.0±0.4 | 95.2±1.9 |
| | Sreeja 2021 [23] | 19 | 2D | U-net | 67.5±17.3 | 0.86±0.05 | - | - | - | - |
| | Tang 2021 [24] | 27 | 2D | GAN | 60.8±14.0 | - | 49.23±1.92 | 99.96 | 98.0 | - |
| | Zimmermann 2021 [10][f] | 33 | 3D | U-net | 68.1±5.4 | 0.97±0.00 | - | - | - | - |
| | Gholamiankhah 2022[25] | 86 | 2D | CNN | 114±27.5 | 0.95±0.04 | 28.7±1.59 | - | - | - |
| | Wang 2022[11][g] | 145 | 2D | GAN | 50.2±18 | 0.92±0.03 | 31.8±2.6 | - | - | - |
| T1wGd | Emami 2018 [9] and Liu, 2021 [26][h] | 15 | 2D | GAN | 89.3±10.3 | 0.83±0.03 | 26.6±1.2 | - | 99.9±0.2 | 99.0±1.5 |
| | Kazemifar 2019 [27] | 63 | 2D | GAN | 47.2±11.0 | - | - | - | 99.2±0.8 | 94.6±2.9 |
| | Liu, 2019 [28] | 40 | 2D | GAN | 75±23 | - | - | 99.2 | - | - |
| | Alvarez-Andres 2020 [22][f] | 133 | 3Dp | CNN | 87±28 | - | - | 99.9±0.18 | 99.6±0.30 | 97.9±1.07 |
| | Massa 2020 [6] | 81 | 2D | U-net | 44.6±7.48 | 0.64±0.03 | 43.4±1.22 | - | - | - |
| | Li 2021 [29] | 18 | 2D | GAN | 74.9±15.6 | 0.83±0.04 | 27.7±1.43 | - | - | - |
| | Zimmermann 2021 [10][f] | 24 | 3D | U-net | 71.6±9.4 | 0.96±0.01 | - | - | - | - |

The table continues on the next page.

Table 10: Literature review: brain sCT with image similarity metrics and gamma analysis (Continued).

| Training sequence | Reference | Pts[a] | Conf.[b] | Model | Image similarity ||| Gamma analysis |||
|---|---|---|---|---|---|---|---|---|---|---|
| | | | | | MAE [HU] | SSIM | PSNR [dB] | $\gamma_{3\%,3mm}$ [%] | $\gamma_{2\%,2mm}$ [%] | $\gamma_{1\%,1mm}$ [%] |
| T1w + T1wGd | Alvarez-Andres 2020 [22][f] | 242 | 3Dp | CNN | 81±22 | - | - | 99.8±0.19 | 99.6±0.32 | 97.9±1.06 |
| | Maspero 2020 [30][g] | 40 | 2D+[e] | GAN | 61.0±14.1 | - | 26.7±1.9 | 99.7±0.6 | 99.6±1.1 | - |
| | Jabbarpour 2022 [12][i] | 60 | 2D | GAN | 62.7±30.7 | 0.88±0.05 | 27.0±3.38 | 99.0±1.10 | 95.0±3.68 | 90.1±6.05 |
| T2wGd | Li 2020 [31] | 28 | 2D | U-net | 65.4±4.08 | 0.97±0.004 | 28.84±0.57 | - | - | - |
| | Massa 2020 [6] | 81 | 2D | U-net | 45.7±8.78 | 0.63±0.03 | 43.4±1.18 | - | - | - |
| | Ranjan 2021 [32] | 18 | 2D | GAN | 0.03±0.02[j] | 0.82±0.06 | 21.4±3.96 | - | - | - |
| | Zimmermann 2021 [10][f] | 32 | 3D | U-net | 68.3±7.3 | 0.98±0.00 | - | - | - | - |
| | Wang 2022[11][g] | 145 | 2D | GAN | 53.7±21 | 0.91±0.02 | 32.5±2.2 | - | - | - |
| T1w + T1wGd + T2w | Zimmermann 2021 [10][f] | 33 | 3D | U-net | T1w: 69.1±5.7 T1wGd: 70.0±8.4 T2w: 67.3±7.1 | T1w: 0.97±0.00 T1wGd: 0.97±0.01 T2w: 0.98±0.00 | - | - | - | - |
| FLAIR | Alvarez-Andres 2020 [22][f] | 134 | 3Dp | CNN | 115±22 | - | - | - | - | - |
| | Massa 2020 [6] | 81 | 2D | U-net | 51.2±4.5 | 0.61±0.04 | 44.9±1.15 | - | - | - |
| | Wang 2022[11][g] | 145 | 2D | GAN | 59.3±22 | 0.91±0.03 | 31.3±2.0 | - | - | - |
| T2w + FLAIR | Jabbarpour 2022 [12][i] | 65 | 2D | GAN | 61.9±22.6 | 0.84±0.05 | 27.1±2.25 | - | - | - |
| Multichannel: T1w + T2w + FLAIR | Koike 2019 [17] | 15 | 2Dp | GAN | 108±24.0 | - | - | 99.8±0.3 | 99.2±1.0 | 95.3±4.7 |

For references where multiple models were compared, results are only reported for the best performing model unless models were trained for different MRI sequences. [a]Number of patients in the training set. [b]Configuration, p: patch-based training. [c]Three orthogonal slices as input. [d]Three consecutive slices as input. [e]Combined output from three networks that take one direction from the three orthogonal planes as input. [f]Results reported for several single-sequence models and a combined model. [g]Paediatric population. [h]Dosimetry reported in [26] for the model presented in [9]. [i]Heterogeneous imaging protocol, incl. images +/- Gd and +/- FLAIR. [j]MAE not in HU.